\documentclass[useAMS, twocolumn]{mn2e}
\usepackage{graphicx}
\usepackage{amssymb}
\usepackage{amsmath}

\newcommand\be{\begin{equation}}
\newcommand\ee{\end{equation}}

\title[Circumbinary Accretion Discs]{Viscous Hydrodynamics Simulations of Circumbinary Accretion Discs: Variability, Quasi-Steady State, and Angular Momentum Transfer}
\author[R. Miranda, D. Mu{\~n}oz, and D. Lai]{Ryan Miranda$^1$\thanks{rjm456@cornell.edu}, Diego J. Mu{\~n}oz$^{1,2}$, and Dong Lai$^{1,3}$ \\
                                                                        $^1$Cornell Center for Astrophysics and Planetary Science, Department of Astronomy, Cornell University, Ithaca, NY 14853, USA \\
                                                                        $^2$Physics Department, Technion - Israel Institute of Technology, Haifa, Israel 3200003 \\
                                                                        $^3$Institute for Advanced Study, Princeton, NJ 08540, USA}

\begin{document}

\maketitle

\begin{abstract}
We carry out numerical simulations of circumbinary discs, solving the viscous hydrodynamics equations on a polar grid covering an extended disc outside the binary co-orbital region. We use carefully controlled outer boundary conditions and long-term integrations to ensure that the disc reaches a quasi-steady state, in which the time-averaged mass accretion rate onto the binary, $\langle\dot{M}\rangle$, matches the mass supply rate at the outer disc. We focus on binaries with comparable masses and a wide range of eccentricities ($e_\mathrm{B}$). For $e_\mathrm{B} \lesssim 0.05$, the mass accretion rate of the binary is modulated at about $5$ times the binary period; otherwise it is modulated at the binary period. The inner part of the circumbinary disc ($r \lesssim 6 a_\mathrm{B}$) generally becomes coherently eccentric. For low and high $e_\mathrm{B}$, the disc line of apsides precesses around the binary, but for intermediate $e_\mathrm{B}$ ($0.2 - 0.4$), it instead becomes locked with that of the binary. By considering the balance of angular momentum transport through the disc by advection, viscous stress, and gravitational torque, we determine the time-averaged net angular momentum transfer rate to the binary, $\langle\dot{J}\rangle$. The specific angular momentum, $l_0 = \langle\dot{J}\rangle/\langle\dot{M}\rangle$, depends non-monotonically on $e_\mathrm{B}$. Contrary to previous claims, we find that $l_0$ is positive for most $e_\mathrm{B}$, implying that the binary receives net angular momentum, which may cause its separation to grow with time. The minimum $l_0$ occurs at intermediate $e_\mathrm{B}$ ($0.2 - 0.4$), corresponding to the regime where the inner eccentric disc is apsidally aligned with the binary.
\end{abstract}


\section{Introduction}
\label{sec:introduction}

Circumbinary discs have been observed in a number of young stellar binaries (e.g., Dutrey et al.~1994; Mathieu et al.~1997; Simon et al.~2000). They are also expected to exist around supermassive black hole (SMBH) binaries as a consequence of accretion from the interstellar medium following galaxy mergers (e.g., Begelman et al.~1980; Ivanov et al. 1999; Milosavljevi{\'c} \& Phinney 2005; Escala et al.~2005; Mayer et al.~2007; Dotti et al.~2007; Cuadra et al.~2009; Chapon et al.~2013). Understanding the dynamical behavior of circumbinary discs has important applications for a variety of astrophysical problems, including variable accretion in both young stellar objects (Jensen et al. 2007; Muzerolle et al. 2013; Bary \& Petersen 2014) and active galactic nuclei (e.g., D'Orazio et al. 2015), circumbinary planet formation (Paardekooper et al. 2012; Meschiari 2012; Rafikov 2013; Silsbee \& Rafikov 2015), and long-term binary orbital evolution (e.g., Armitage \& Natarajan 2002; Haiman et al.~2009; Chang et al. 2010; Roedig et al.~2012). In this paper we carry out a suite of numerical simulations to investigate the structure, morphology, and variability of circumbinary accretion on a wide range of time-scales, as well as the long-term angular momentum transfer between the binary and the circumbinary disc.

Several characteristic features distinguish circumbinary discs from circumsingle discs. A central low-density cavity, with a size of a few times larger than the binary orbit, is carved out around the binary by resonant gravitational torques, which prevent direct viscous inflow of mass (e.g., Artymowicz \& Lubow 1994; Miranda \& Lai 2015; note that binaries with small mass ratios open annular gaps, rather than common cavities, see D'Orazio et al. 2016). However, accretion of mass by the binary is by no means suppressed, and proceeds via narrow, non-axisymmetric streams which penetrate the central cavity (e.g., Artymowicz \& Lubow 1996; MacFadyen \& Milosavljevi\'{c}  2008; Mu\~{n}oz \& Lai 2016). These lead to time variability of the binary mass accretion rate. Streams of material may also be launched outward into the cavity by orbital motion of the binary (Mu{\~n}oz \& Lai 2016). Additionally, the binary excites eccentricity in the disc, possibly due to eccentric Lindblad resonances (Lubow 1991), and the eccentric disc may precess around the binary (e.g., Nelson 2003; Shi et al. 2012), potentially producing long-term variabilities (Mu\~{n}oz \& Lai 2016).

The variable mass accretion of the binary on orbital time-scales has been demonstrated both numerically (e.g., MacFadyen \& Milosavljevi\'{c}  2008; Mu\~{n}oz \& Lai 2016) and observationally (Jensen et al. 2007; Muzerolle et al. 2013; Bary \& Petersen 2014). Numerically, the accretion rate is found to vary primarily with a period equal to either the binary orbital period ($P_\mathrm{B}$), or about $5$ times the binary orbital period. The occurrence of the longer period ($5 P_\mathrm{B}$) variability is found to be linked to the properties of the binary. For example, it is suppressed for circular binaries with sufficiently small mass ratios (Farris et al. 2014). It is also suppressed when the binary is eccentric (Mu\~{n}oz \& Lai 2016). However, the detailed dependence of this behavior on the binary eccentricity has not been fully explored.

There have been conflicting claims in the literature about how the average mass accretion rate of the binary compares to that of a single star of the same mass. The ratio of the former to the latter characterizes the extent to which gravitational torques from the binary suppress or enhance accretion. Previous studies have reported values of this ratio both smaller than unity (e.g., MacFadyen \& Milosavljevi{\'c} 2008; Ragusa et al. 2016), and greater than unity (e.g., Farris et al. 2014). In considering these claims, it is necessary to make a distinction between discs whose outer edges spread freely, and those whose outer regions are supplied with mass at a steady rate (e.g., due to an infalling envelope). For a freely spreading disc, a steady state, in which the accretion rate is constant throughout the disc, is never reached, and so the accretion rate of the binary continually changes with time (e.g., Rafikov 2016). For a disc with a steady mass supply at the outer region, a quasi-steady state is possible. In such a state, the time-averaged accretion rate of the binary must necessarily be equal to the rate at which mass is supplied to the outer disc, which is governed by processes occurring far from the binary. The time-averaged accretion rate of the central object must then be independent of whether it is a single body or a binary. The quasi-steady state was demonstrated in the simulations of Mu\~{n}oz \& Lai (2016) using the moving mesh code \textsc{arepo}, which traces the gas from a circumbinary disc via accretion streams to circumsingle discs.

The rate at which angular momentum is transferred to the binary from the circumbinary disc is an outstanding open question. If there were no mass accretion onto the binary, the angular momentum coupling would be mediated only through gravitational torque, and the binary would lose angular momentum to the surrounding disc. It is generally thought that such angular momentum loss could play a central role in the mergers of SMBH binaries--shrinking their orbits from $\sim 1 \mathrm{pc}$ to $\sim 0.01 \mathrm{pc}$ separations, at which gravitational radiation will cause them to merge within a Hubble time--providing a possible solution to the so-called ``final parsec problem''  (e.g., Begelman et al. 1980, Armitage \& Natarajan 2002; Wyithe \& Loeb 2003; Jaffe \& Backer 2003; Sesana et al. 2008; MacFadyen \& Milosavljevi{\'c} 2008; Haiman et al.~2009; Kelley et al. 2016). However, mass accretion carries positive angular momentum to the binary (in the case of a prograde disc). As a result of the competition between these two effects, the binary will, on average, either experience a net gain or suffer a net loss of angular momentum. The sign of this net angular momentum transfer has important consequences for the structure of the circumbinary disc, and directly determines (along with the rate of orbital energy transfer) the orbital evolution of the binary. As circumbinary accretion is generally variable on a range of time-scales, controlled, long-duration numerical simulations are required to quantify the balance between gravitational torques and angular momentum advection.

In this paper, we present a series of 2D viscous hydrodynamics simulations of circumbinary accretion. While we adopt the simplest physical ingredients in our simulations (e.g., no magnetic fields or radiation transfer are included, and viscosity is prescribed using the $\alpha$ ansatz), our main goal is to carry out numerical experiments systematically in a well-controlled manner and for sufficiently long durations, so that reliable information can be obtained for the long-term evolution of the disc and for the net angular momentum transfer rate to the binary. To this end, we feed the outer disc boundary with a constant mass supply and ensure that the outer disc is close to the steady state. We carry out simulations over many viscous times of the ``mid-disc'' region (far away from the disc truncation radius), and ensure that a large region of the disc reaches a quasi-steady state. We survey a wide range of binary eccentricity, and to a lesser extent, the binary mass ratio and disc viscosity parameter, and study how the disc behavior and evolution depends on these parameters. We find that, in addition to short-term variabilities of the binary accretion rate, the disc also exhibits long-term variabilities, corresponding to the coherent precession and apsidal locking of the inner disc. On average, the central binary can always accept mass at a rate very close to that supplied from the outer disc. Most importantly, we find that the net angular momentum received by the binary, including contributions from mass accretion, viscosity, and gravitational torque, depends on the binary eccentricity ($e_\mathrm{B}$), and is positive for a wide range of $e_\mathrm{B}$. This dependence of the angular momentum transfer on $e_\mathrm{B}$ correlates with the secular behaviors of the inner eccentric disc. Thus, contrary to the widely-held presumption that the binary loses angular momentum to the circumbinary disc, we find that when the effect of mass accretion is accounted for, the binary generally gains angular momentum to the extent that its semi-major axis tends to grow with time.

The plan for this paper is as follows. In Section \ref{sec:setup}, we describe our numerical setup. In Section \ref{sec:morphology}, we examine the morphological features of the disc, and discuss the truncation of the inner cavity. Section \ref{sec:short_timescale} examines the variability of the mass accretion rate of the central binary on short (orbital) time-scales. In Section \ref{sec:eccentricity_precession}, we present numerical results on the long-term variation of the disc; such variation manifests as the precession and apsidal locking of the inner disc. In Section \ref{sec:eccentricity_precession_theory}, which is self-contained, we explore possible theoretical explanations for the apsidal locking phenomenon. In Section \ref{sec:longterm_mdot_jdot}, we investigate the long-term evolution of the global mass accretion rate and angular momentum accretion rate of the disc, and determine the average rate at which angular momentum is transferred to the binary. We summarize and discuss our results in Section \ref{sec:discussion}.

\section{Problem Setup and Overview}
\label{sec:setup}

A variety of computational approaches have been used to investigate the problem of circumbinary accretion, including both smoothed-particle hydrodynamics (e.g., Artymowicz \& Lubow 1996; Cuadra et al. 2009; Roedig et al. 2012; Pelupessy \& Portegies Zwart 2013; Dunhill et al. 2015) and Eulerian methods. There is a dichotomy in the Eulerian approaches between those which include the co-orbital region of the binary in the computational domain, and those which excise this region. When the co-orbital region is excluded (e.g., MacFadyen \& Milosavljevi\'{c}  2008; Shi et al. 2012; D'Orazio et al. 2013; Shi \& Krolik 2015; Lines et al. 2015), the flow can be solved efficiently using a polar grid geometry and an orbital advection algorithm (e.g., FARGO; Masset 2000) to subtract out the largely azimuthal average fluid motion, due to the fact that it is largely azimuthal. When the co-orbital region, where the flow is much less uniform, is included, the efficiency of this approach is lost. In this case, other methods, including the use of Cartesian grids (e.g., G\"{u}nther \& Kley 2002; Hanawa et al. 2010; de Val-Borro et al. 2011) or moving meshes (e.g., Farris et al. 2014; Mu\~{n}oz \& Lai 2016), have been adopted. In this paper, we employ the polar grid method with an excised binary co-orbital region. Although we do not follow the details of the flow around and onto the individual members of the binary, we take advantage of the lower computational cost in order to perform long-term (viscous time-scale) integrations, and to explore a wide variety of binary orbital parameters. 

\subsection{Numerical Setup}

We solve the viscous hydrodynamic equations describing a thin disc around a binary, which consists of masses $M_1$ and $M_2$ (total mass $M_\mathrm{B} = M_1 + M_2$), with mass ratio $q_\mathrm{B} = M_2/M_1$, semi-major axis $a_\mathrm{B}$, and eccentricity $e_\mathrm{B}$. The positions ($r_i, \phi_i$ in polar coordinates) of $M_1$ and $M_2$ as a function of time are obtained by solving Kepler's equation using a higher-order extension of Newton's method (e.g., Murray \& Dermott 1999). In our simulations, the circumbinary disc extends from $r_\mathrm{in} = (1+e_\mathrm{B})a_\mathrm{B}$ to $r_\mathrm{out} = 70 a_\mathrm{B}$, and is subject to the total gravitational potential of the binary,
\be
\Phi(r,\phi,t) = -\sum_{i=1}^2\frac{GM_i}{\left[r^2 + r_i^2 - 2rr_i\cos\left(\phi-\phi_i\right)\right]^{1/2}}.
\ee
The equation of state is locally isothermal,
\be
P = c_\mathrm{s}^2(r)\Sigma,
\ee
where $c_\mathrm{s}(r) = h r\Omega_\mathrm{K}$ is the sound speed and $\Omega_\mathrm{K} = (GM_\mathrm{B}/r^3)^{1/2}$. The disc aspect ratio, $h = H/r$ (where $H = c_\mathrm{s}/\Omega_\mathrm{K}$ is the pressure scale height) is therefore constant with $r$. Throughout this paper we choose $h = 0.1$. The kinematic viscosity is prescribed using the $\alpha$-ansatz, $\nu = \alpha H c_\mathrm{s} = \alpha h^2 r^2\Omega_\mathrm{K}$.

The initial surface density of the disc is
\be
\label{eq:sigma_t0}
\Sigma(r) = \frac{\dot{M}_0}{3\pi\alpha h^2\sqrt{GM_\mathrm{B}r}} \left[1 - \left(\frac{r_\mathrm{in}}{r}\right)^{1/2}\right] \exp\left[-\left(\frac{r}{r_\mathrm{edge}}\right)^{-2}\right], 
\ee
and the initial radial velocity is
\be
\label{eq:ur_t0}
u_r(r) = -\frac{3\nu}{2r} \left[1 - \left(\frac{r_\mathrm{in}}{r}\right)^{1/2}\right]^{-1}.
\ee
For $r \gg r_\mathrm{edge}$, these profiles correspond to a steady state, constant $\dot{M}$ disc, with a zero torque condition at $r_\mathrm{in}$. The Gaussian factor in Eq.~(\ref{eq:sigma_t0}) creates an artificial cavity around the central binary at the approximate radius of the real cavity which is eventually self-consistently produced, in order to avoid violent relaxation at the beginning of the simulation. We set $r_\mathrm{edge} = 2 r_\mathrm{in}$. The initial rotation profile of the disc is in centrifugal balance, including contributions from the binary quadrupole potential and pressure gradients,
\be
\Omega^2(r) = \Omega_\mathrm{K}^2\left[1 + \frac{3}{4}\frac{q_\mathrm{B}}{(1+q_\mathrm{B})^2}\left(1 + \frac{3}{2}e_\mathrm{B}^2\right)\left(\frac{r}{a_\mathrm{B}}\right)^{-2}\right] + \frac{1}{r\Sigma}\frac{\mathrm{d}P}{\mathrm{d}r}.
\ee

At the outer boundary $r_\mathrm{out}$, all fluid variables are assumed to be fixed at their steady state values, and mass is injected into the domain at a rate $\dot{M}_0$. In code units ($GM_\mathrm{B} = a_\mathrm{B} = \Omega_\mathrm{B} = 1$), we choose $\dot{M}_0 = 3\pi\alpha h^2$, so that $\Sigma \approx r^{-1/2}(1-\sqrt{r_\mathrm{in}/r})$ at $r \sim r_\mathrm{out}$. At the inner boundary, we employ a ``diode'' boundary condition, in which zero-gradient conditions ($\partial/\partial r = 0$) are imposed on $\Sigma$ and $u_\phi$, as well as on $u_r$, whenever it is negative. When $u_r$ is positive, it is instead reflected across the boundary so that $u_r(r_\mathrm{in}) = 0$. This ensures that mass is allowed to leave the domain but cannot enter it through the inner boundary.

The fluid equations are solved using the finite volume, shock-capturing hydrodynamics code \textsc{pluto} (Mignone et al. 2007). We use third-order Runge-Kutta time stepping, piecewise parabolic spatial reconstruction, a Roe method Riemann solver, and the FARGO orbital advection algorithm. We use a polar grid centred on the centre of mass of the binary, with uniform grid spacing in the azimuthal direction and logarithmic grid spacing in the radial direction. Unless otherwise stated, $N_\phi = 600$ azimuthal grid cells are used. The number of radial cells is chosen so that $\Delta r \approx r\Delta\phi$, i.e., the cells are approximately square, by setting $N_r \approx [N_\phi/(2\pi)]\ln(r_\mathrm{out}/r_\mathrm{in})$. All runs have the same value of $r_\mathrm{out} (= 70 a_\mathrm{B})$, but different values of $r_\mathrm{in} = (1 + e_\mathrm{B}) a_\mathrm{B}$, and therefore different values of $N_r$.

\subsection{Analysis Procedure}

\begin{table*}
\begin{center}
\begin{tabular}{|c|c|c|c|c|c|c|c|c|c|}

\hline

$q_\mathrm{B}$ & $e_\mathrm{B}$ & $\alpha$ & $t_\mathrm{end}$ $[P_\mathrm{B}]$ & $r_\mathrm{rel}(t_\mathrm{end})$ $[a_\mathrm{B}]$ & $r_\mathrm{peak}$ $[a_\mathrm{B}]$ & $r_{0.1}$ $[a_\mathrm{B}]$ & Lump & $\dot{\varpi}_\mathrm{d}$ $[\Omega_\mathrm{B}]$ & $l_0$ $[a_\mathrm{B}^2\Omega_\mathrm{B}]$ \\ \hline

$1.0$ & $0.0$  & $0.1$   & $3000$  & $12.16$ & $3.56$ & $1.67$ & Yes & $3.38 \times 10^{-3}$ & $+0.816 \pm 0.027$ \\
$1.0$ & $0.05$ & $0.1$   & $3000$  & $12.16$ & $3.52$ & $1.66$ & Yes & $3.32 \times 10^{-3}$ & $+0.825 \pm 0.025$    \\
$1.0$ & $0.1$  & $0.1$   & $3000$  & $12.16$ & $3.64$ & $1.65$ & No  & $7.64 \times 10^{-4}$ & $+0.690 \pm 0.098$ \\
$1.0$ & $0.2$  & $0.1$   & $5000$  & $17.10$ & $3.40$ & $1.80$ & No  &  None                 & $+0.674 \pm 0.021$ \\
$1.0$ & $0.4$  & $0.1$   & $8000$  & $23.39$ & $3.15$ & $2.02$ & No  &  None                 & $+0.273 \pm 0.017$ \\
$1.0$ & $0.6$  & $0.1$   & $5000$  & $17.10$ & $4.58$ & $2.23$ & No  & $1.92 \times 10^{-3}$ & $+1.053 \pm 0.018$ \\
$1.0$ & $0.8$  & $0.1$   & $4000$  & $14.73$ & $5.27$ & $2.34$ & No  & $1.79 \times 10^{-3}$ & $+1.249 \pm 0.012$ \\
$1.0$ & $0.0$  & $0.05$  & $7000$  & $13.48$ & $3.95$ & $1.88$ & Yes & $2.71 \times 10^{-3}$ & $+0.800 \pm 0.061$ \\
$1.0$ & $0.2$  & $0.05$  & $25000$ & $31.49$ & $3.13$ & $1.95$ & No  & None                  & $-0.481 \pm 0.145$ \\
$1.0$ & $0.3$  & $0.05$  & $10000$ & $17.10$ & $4.63$ & $2.18$ & No  & $1.44 \times 10^{-3}$ & $+0.996 \pm 0.056$ \\
$1.0$ & $0.4$  & $0.05$  & $12000$ & $19.31$ & $4.49$ & $2.28$ & No  & $1.80 \times 10^{-3}$ & $+0.971 \pm 0.053$ \\
$1.0$ & $0.8$  & $0.05$  & $10000$ & $17.10$ & $5.33$ & $2.57$ & No  & $1.60 \times 10^{-3}$ & $+1.196 \pm 0.032$ \\
$0.5$ & $0.0$  & $0.1$   & $3000$  & $12.16$ & $3.45$ & $1.63$ & Yes & $3.18 \times 10^{-3}$ & $+0.830 \pm 0.043$ \\
$0.5$ & $0.4$  & $0.1$   & $5000$  & $17.10$ & $4.35$ & $1.96$ & No  & $1.44 \times 10^{-3}$ & $+1.030 \pm 0.028$ \\
$0.2$ & $0.0$  & $0.1$   & $3000$  & $12.16$ & $3.08$ & $1.42$ & No  & $3.33 \times 10^{-3}$ & $+0.873 \pm 0.020$ \\
$0.2$ & $0.4$  & $0.1$   & $3000$  & $12.16$ & $4.68$ & $1.69$ & No  & None                  & $+1.119 \pm 0.054$ \\

\end{tabular}
\end{center}
\caption{Summary of parameters and key results of all simulations presented in this paper. The first five columns give the binary mass ratio $q_\mathrm{B}$ and eccentricity $e_\mathrm{B}$, the disc viscosity parameter $\alpha$, the total integration time $t_\mathrm{end}$, and the viscous relaxation radius at $t_\mathrm{end}$. The next two columns are the two radii characterizing the inner disc truncation, $r_\mathrm{peak}$ and $r_{0.1}$. The column labelled ``Lump'' indicates the presence or absence of a lump feature which causes variability of the mass accretion rate with a period of $\sim 5 P_\mathrm{B}$. The second to last column gives the precession frequency of the eccentric inner disc, or indicates if it is instead apsidally aligned with the binary (if ``None'' is listed). The last column is the net angular momentum received by the binary per unit accreted mass.}
\label{tab:summary}
\end{table*}

Table \ref{tab:summary} summarizes the parameters and key results of all the simulations presented in this paper. The relevant quantities and results are discussed in the main sections of the paper. Of primary interest to our study is allowing the disc to reach a steady state. This requires that the influence of the central binary be communicated, by viscosity, to a sufficiently large radius. The viscous time-scale at $r$ is $t_\nu = (4/9)(r^2/\nu)$ (Lynden-Bell \& Pringle 1974). Therefore, after a time $t$, the disc is viscously relaxed within a radius $r_\mathrm{rel}$, defined by $t = t_\nu(r_\mathrm{rel})$:
\be
\label{eq:r_relax}
r_\mathrm{rel}(t) = \left(\frac{9\pi}{2}\alpha h^2\frac{t}{P_\mathrm{B}}\right)^{2/3} a_\mathrm{B}.
\ee
Table \ref{tab:summary} indicates the total integration time $t_\mathrm{end}$ of each simulation, and its corresponding viscous relaxation radius $r_\mathrm{rel}(t_\mathrm{end})$. In all cases, we allow the disc to relax out to at least $12 a_\mathrm{B}$, which is significantly larger than the inner disc radius. However, this is only a necessary, not a sufficient condition, for the disc to be in a quasi-steady state. In Section \ref{sec:longterm_mdot_jdot}, we discuss additional criteria we use to determine that the disc has fully relaxed.

During the long-term evolution ($t = [0,t_\mathrm{end}]$) of each simulation, the fluid variables are output every $10 P_\mathrm{B}$. This information is used to determine the secular evolution of the orbital elements of the disc, in particular, whether the disc precesses or is apsidally aligned with the binary (see Section \ref{sec:eccentricity_precession}). Several shorter duration runs are restarted from snapshots taken during the long-term evolution. The duration of the shorter runs are chosen according to the long-term behavior of the orbital elements of the disc. For precessing discs, the duration is equal to the precession period (as this is the longest period over which variability occurs), to the nearest $50 P_\mathrm{B}$, and the fluid variables are output $5$ times per $P_\mathrm{B}$. For the aligned cases, the restarted runs last only $50 P_\mathrm{B}$, and the variables are output $20$ times per $P_\mathrm{B}$. In either case, at least $1000$, and as many as $3500$ snapshots of the fluid variables are captured in each restart run. The snapshots are used to compute time-averaged profiles of the mass accretion rate $\dot{M}(r)$, and the net angular momentum accretion rate $\dot{J}(r)$ (see Section \ref{sec:longterm_mdot_jdot} and Eq.~\ref{eq:jdotdef}). We use the evolution of these quantities to verify that the simulations are evolving towards a quasi-steady state, as described in Section \ref{sec:longterm_mdot_jdot}. Finally, each simulation is restarted from $t_\mathrm{end}$, and evolved for an additional $50 P_\mathrm{B}$, with a sampling rate of $20/P_\mathrm{B}$. This restart run is used to compute the time-averaged surface density profiles shown in Section \ref{sec:morphology}, as well as the time series of $\dot{M}$ shown in Section \ref{sec:short_timescale}. These analyses are therefore performed while the disc has reached the quasi-steady state.

In order to check the effects of numerical resolution, we also performed two higher resolution runs, with $N_r$ and $N_\phi$ both twice as large as in the standard runs, for the cases $(q_\mathrm{B}, e_\mathrm{B}, \alpha) = (1.0,0.4,0.1)$ and $(1.0,0.2,0.05)$. Due to their high computational cost, these were only evolved for $1000 P_\mathrm{B}$, followed by the $50 P_\mathrm{B}$ restart runs. We verified that the features and behaviors found in these runs were largely unchanged compared to their standard resolution counterparts.

\section{Disc Morphology}
\label{sec:morphology}

\begin{figure*}
\begin{center}
\includegraphics[width=0.99\textwidth,clip]{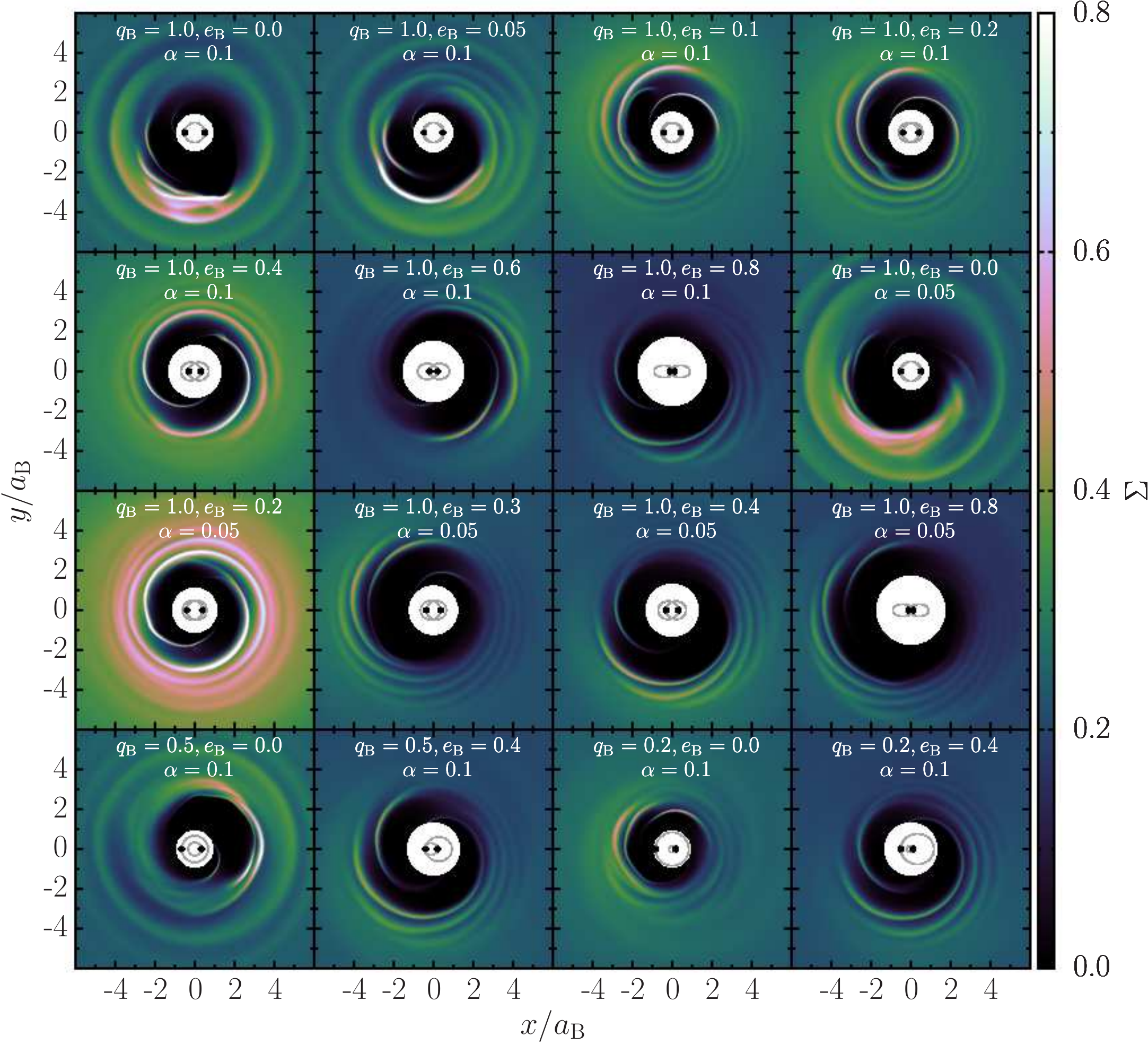}
\caption{Snapshots of surface density in the inner disc after reaching a quasi-steady state (see Section \ref{subsec:general_features}). The snapshot for each simulation is taken at its corresponding time $t_\mathrm{end}$, as given in Table \ref{tab:summary}, at which the binary is always at pericentre. The binary orbit is shown in the centre of each panel. Note that for $e_\mathrm{B} = 0$ (and, to a lesser extent, $e_\mathrm{B} = 0.05$), an $m = 1$ lump develops near the disc inner edge.}
\label{fig:snapshots}
\end{center}
\end{figure*}

\begin{figure}
\begin{center}
\includegraphics[width=0.45\textwidth,clip]{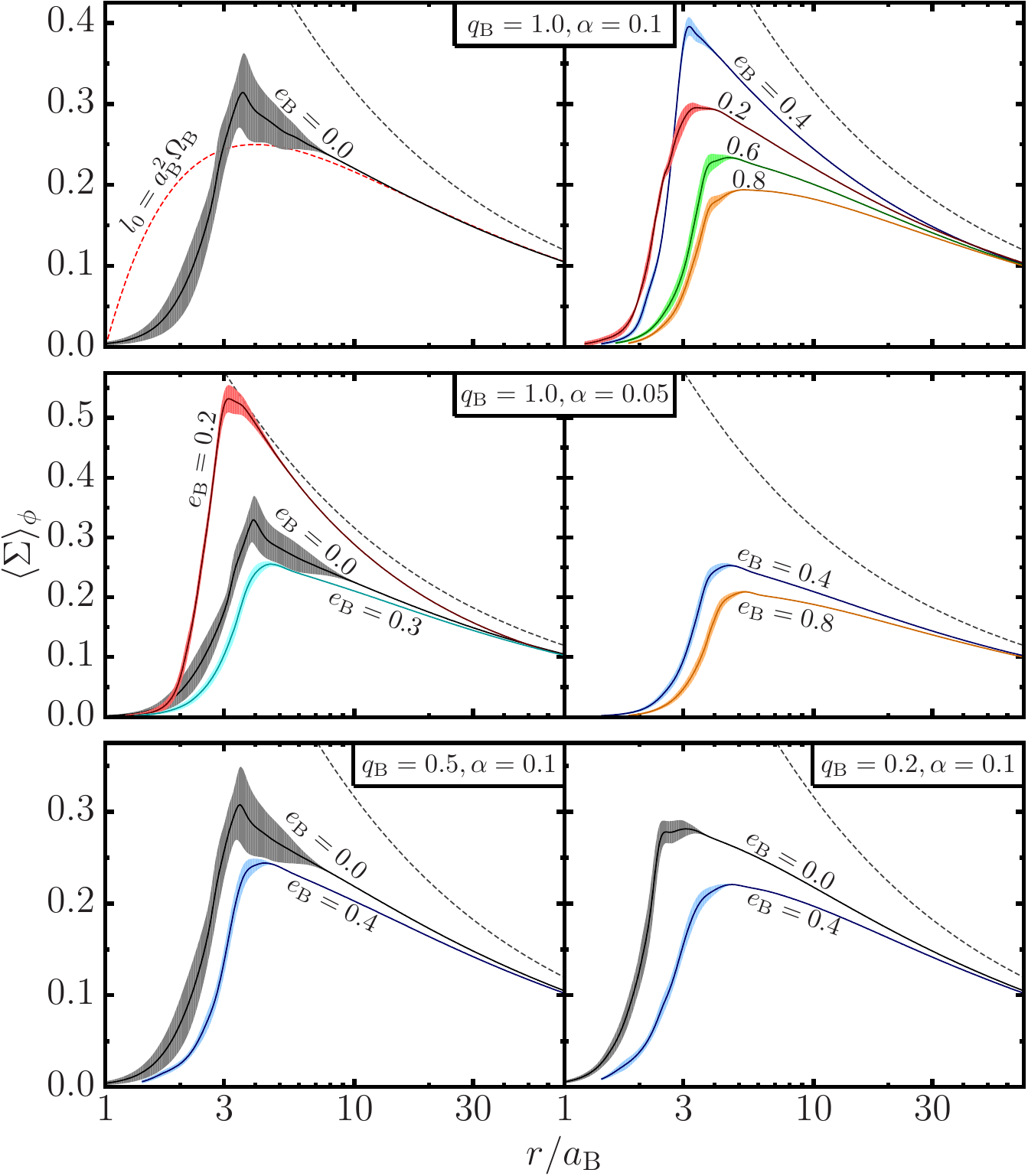}
\caption{Azimuthally-averaged surface density profiles for various binary and disc parameters (see Section \ref{subsec:truncation}). The solid lines are time-averaged over $50$ binary orbits, and the shaded regions around them indicate their $1\sigma$ variations. The black dashed line depicts the asymptotic surface density profile for $\sqrt{r/a_\mathrm{B}} \gg 1$, $\Sigma = \dot{M}_0/(3\pi\nu)$. The red dashed line in the top-left panel shows the deviation from this profile in the outer disc due to the imposed value of $l_0$ (see Eq.~\ref{eq:sigma_t0}) for $e_\mathrm{B} = 0$.}
\label{fig:density_profiles}
\end{center}
\end{figure}

\begin{figure}
\begin{center}
\includegraphics[width=0.45\textwidth,clip]{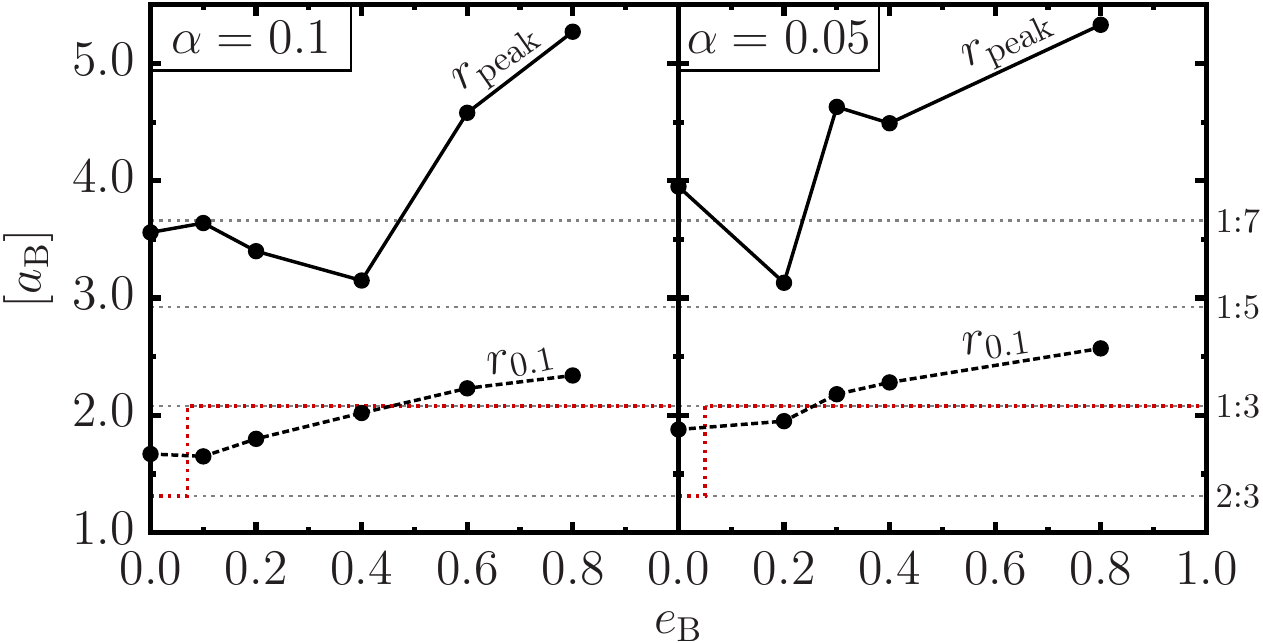}
\caption{The radial locations of two points characterizing the truncation of the inner disc (see Section \ref{subsec:truncation}): $r_\mathrm{peak}$, the point at which the azimuthally averaged surface density is largest, and $r_{0.1}$, where it drops to $10$ per cent of the value at $r_\mathrm{peak}$, as a function of binary eccentricity. Only equal-mass binaries ($q_\mathrm{B} = 1.0$) are shown. The locations of several resonances which are important to disc truncation are shown on the right, and the dashed red line shows the theoretical predictions for the truncation radius as derived by Miranda \& Lai (2015).}
\label{fig:truncation}
\end{center}
\end{figure}

\subsection{General Features}
\label{subsec:general_features}

Figure~\ref{fig:snapshots} shows a sample of surface density snapshots for each of the simulation runs (see Table \ref{tab:summary}). These snapshots are taken near the end of each simulation, when the inner disc has reached a quasi-steady state. Only the innermost part of the disc, where there are strong visible deviations from axisymmetry, is shown. The snapshots illustrate several features that are common to all of the simuations. There is a low-density central cavity, approximately $(2 - 3) a_\mathrm{B}$ in radius, carved out by gravitational torques exerted by the binary (see Section \ref{subsec:truncation}). Spiral density waves, primarily with azimuthal number $m = 1$ or $2$, are excited by the binary and propagate outwards to several times the cavity radius. The central cavity is asymmetric, and is penetrated by dense (relative to the low average surface density in the cavity), narrow accretion streams which carry mass toward the inner boundary. The disc is eccentric, as seen by the shape of the cavity edge (see Section \ref{sec:eccentricity_precession} for a detailed discussion of the eccentricity dynamics of the inner disc). Finally, for circular, or nearly circular ($e_\mathrm{B} < 0.1$) binaries only, there is a overdense lump, i.e., radially localized feature with approximate $m = 1$ symmetry, at the edge of the cavity. The lump orbits the binary with approximately the local Keplerian orbital period, while exhibiting a continuous cycle of creation and destruction with the same period. This process is associated with modulation of the mass accretion rate through the inner boundary at the frequency $\sim\Omega_\mathrm{B}/5$ (see Section \ref{sec:short_timescale}).

\subsection{Inner Disc Truncation}
\label{subsec:truncation}

Figure~\ref{fig:density_profiles} shows the double-averaged (i.e., azimuthally-averaged and time-averaged) surface density profiles for most of the runs depicted in Fig.~\ref{fig:snapshots}. Generically, $\langle\Sigma\rangle$ has a positive slope ($\mathrm{d}\langle\Sigma\rangle/\mathrm{d}r > 0$) in the inner disc (within a few $a_\mathrm{B}$'s from the binary), then reaches a maximum at a few $a_\mathrm{B}$'s and has negative slope beyond that. For some parameters, the maximum manifests as a sharp peak, while for others, it is broader and less well-defined. This is associated with variation in the value of the specific angular momentum eigenvalue $l_0$ for each simulation (see Section \ref{sec:longterm_mdot_jdot}), with smaller values of $l_0$ corresponding to sharper peaks. The asymptotic behavior of $\langle\Sigma\rangle$ for $\sqrt{r/a_\mathrm{B}} \gg 1$ is $\langle\Sigma\rangle = \dot{M}/(3\pi\nu) \propto r^{-1/2}$, which all of the profiles become similar to at large $r$, is also shown in Fig.~\ref{fig:density_profiles}. However, the density profiles do not exactly approach this form, due to the initial ``guess'' for $l_0$ imposed by the surface density profile (Eq.~\ref{eq:sigma_t0}), $l_0^\mathrm{guess} = \sqrt{GM_\mathrm{B} r_\mathrm{in}}$. This is demonstrated in the top left panel, for the case $(q_\mathrm{B},e_\mathrm{B},\alpha) = (1.0, 0.0, 0.1)$. Here the initial profile for the outer disc, characterized by $l_0 = 1$ (in units of $a_\mathrm{B}^2\Omega_\mathrm{B}$) is shown. The actual surface density profile remains nearly identical to this for $r \gtrsim 10 a_\mathrm{B}$, since the influence of the binary has not yet been viscously communicated to this region.

Regardless of its sharpness, $\langle\Sigma\rangle$ always has a maximum, typically located at $r/a_\mathrm{B} \sim 3 - 5$. We denote this radius $r_\mathrm{peak}$. Between $r_\mathrm{peak}$ and the inner boundary of our simulation domain [$r_\mathrm{in} = (1+e_\mathrm{B})a_\mathrm{B}$], $\langle\Sigma\rangle$ drops by about two orders of magnitude, i.e., the disc is truncated interior to $r_\mathrm{peak}$. Specifying the exact location of the ``truncation radius'' is ambiguous, but we define it as the radius (denoted by $r_{0.1}$) at which $\langle\Sigma\rangle$ is $10$ per cent of its value at $r_\mathrm{peak}$.

Figure~\ref{fig:truncation} shows $r_\mathrm{peak}$ and $r_{0.1}$, as a function of binary eccentricity. While $r_\mathrm{peak}$ is not monotonic in $e_\mathrm{B}$ (this is also associated with variations in the value of $l_0$), $r_{0.1}$ strictly increases with $e_\mathrm{B}$. In the theory of inner disc truncation (Artymowicz \& Lubow 1994; Miranda \& Lai 2015), the truncation radius increases with $e_\mathrm{B}$, as the torques applied at increasingly higher order resonances overcome the viscous torque at those locations. The locations of the relevant resonances are shown in Fig.~\ref{fig:truncation}, as well the theoretical prediction for the truncation radius, using the formalism of Miranda \& Lai (2015). The theoretical prediction is very similar for both values of $\alpha$ shown. In both cases, the $\Omega/\Omega_\mathrm{B} = 2/3$ resonance is responsible for disc truncation for $e_\mathrm{B} = 0$, and then, above a small critical eccentricity (which is slightly different for the two values of $\alpha$), the $1/3$ resonance becomes sufficiently strong to truncate the disc. Higher order resonances are never strong enough to truncate the disc for the viscosity parameters ($\alpha = 0.05$ and $0.1$) considered in this paper. This discrete behavior is not observed in our simulations, given the ``fuzziness'' of the disc truncation. The theory predicts that the truncation radius lies between $1.3$ and $2.1 a_\mathrm{B}$, while we find that $r_{0.1}$ is between $1.7$ and $2.6 a_\mathrm{B}$. Thus, there is agreement at the $20$ per cent level between the theory and our numerical results.

\section{Short Time-Scale Variability}
\label{sec:short_timescale}

\begin{figure}
\begin{center}
\includegraphics[width=0.45\textwidth,clip]{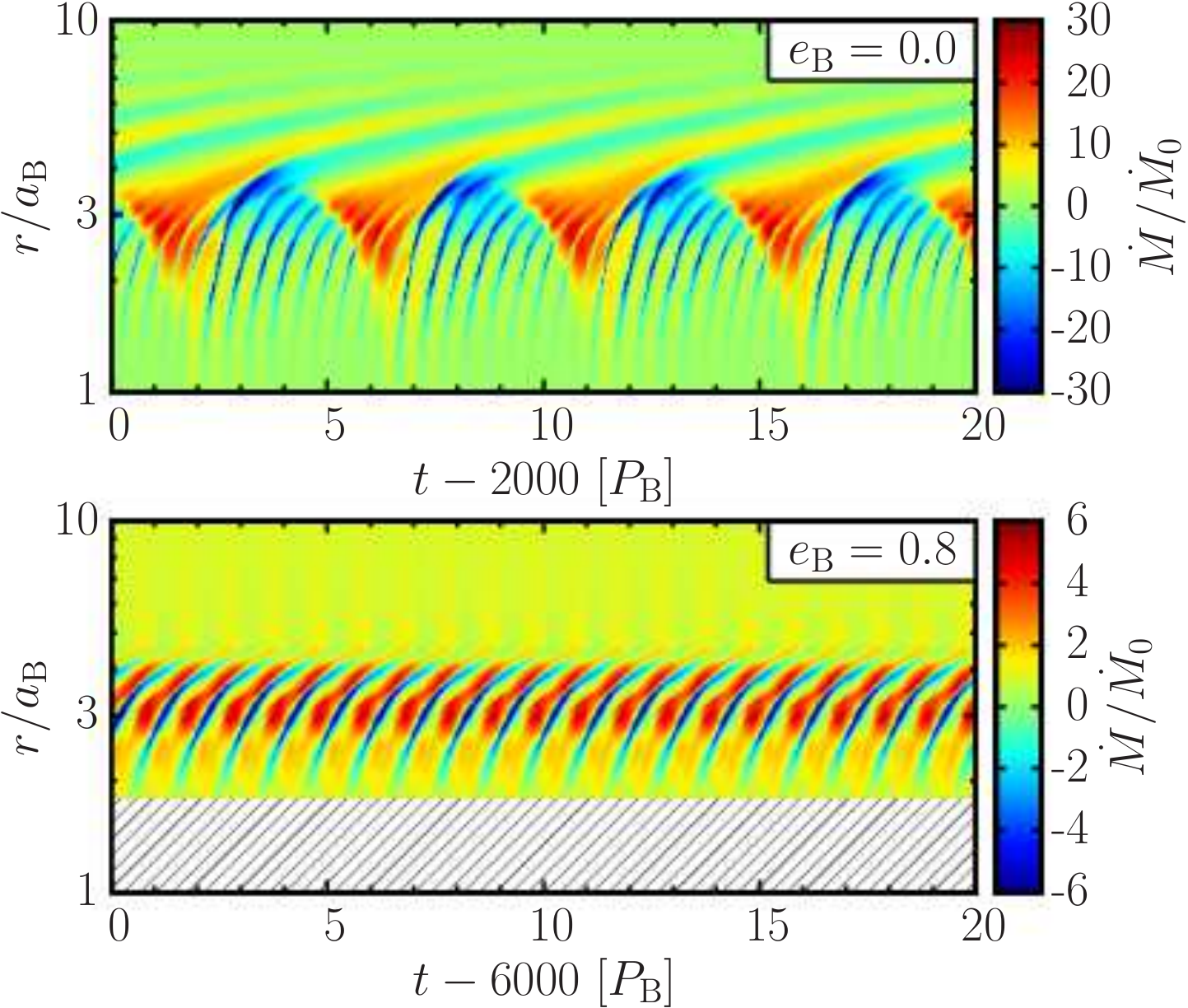}
\caption{The mass accretion rate $\dot{M}(r,t)$ as a function of time ($x$-axis) and radius ($y$-axis), for $q_\mathrm{B}= 1.0$, $\alpha = 0.1$, and two different binary eccentricities, $e_\mathrm{B} = 0.0$ (top) and $0.8$ (bottom). The accretion rate (colours) is normalized by the supplied rate at $r_\mathrm{out}$, $\dot{M}_0$. Note that the colour bars have different scales in the top and bottom panels. The hatched region in the bottom panel indicates that the computational domain does not extend down to $r/a_\mathrm{B} = 1$ as it does in the top panel. For $e_\mathrm{B} = 0.0$, the main periodicities that can be seen are at $5 P_\mathrm{B}$ and $P_\mathrm{B}/2$. For $e_\mathrm{B} = 0.8$, the main variability is at $P_\mathrm{B}$. We find that this behavior is typical for eccentric binaries. In both cases, the accretion rate is steady beyond about $10 a_\mathrm{B}$. This outer region is not shown, in order to focus on the variability of the inner disc (see Section \ref{subsec:mdot_ecc}).}
\label{fig:mdot_r_t}
\end{center}
\end{figure}

\begin{figure}
\begin{center}
\includegraphics[width=0.45\textwidth,clip]{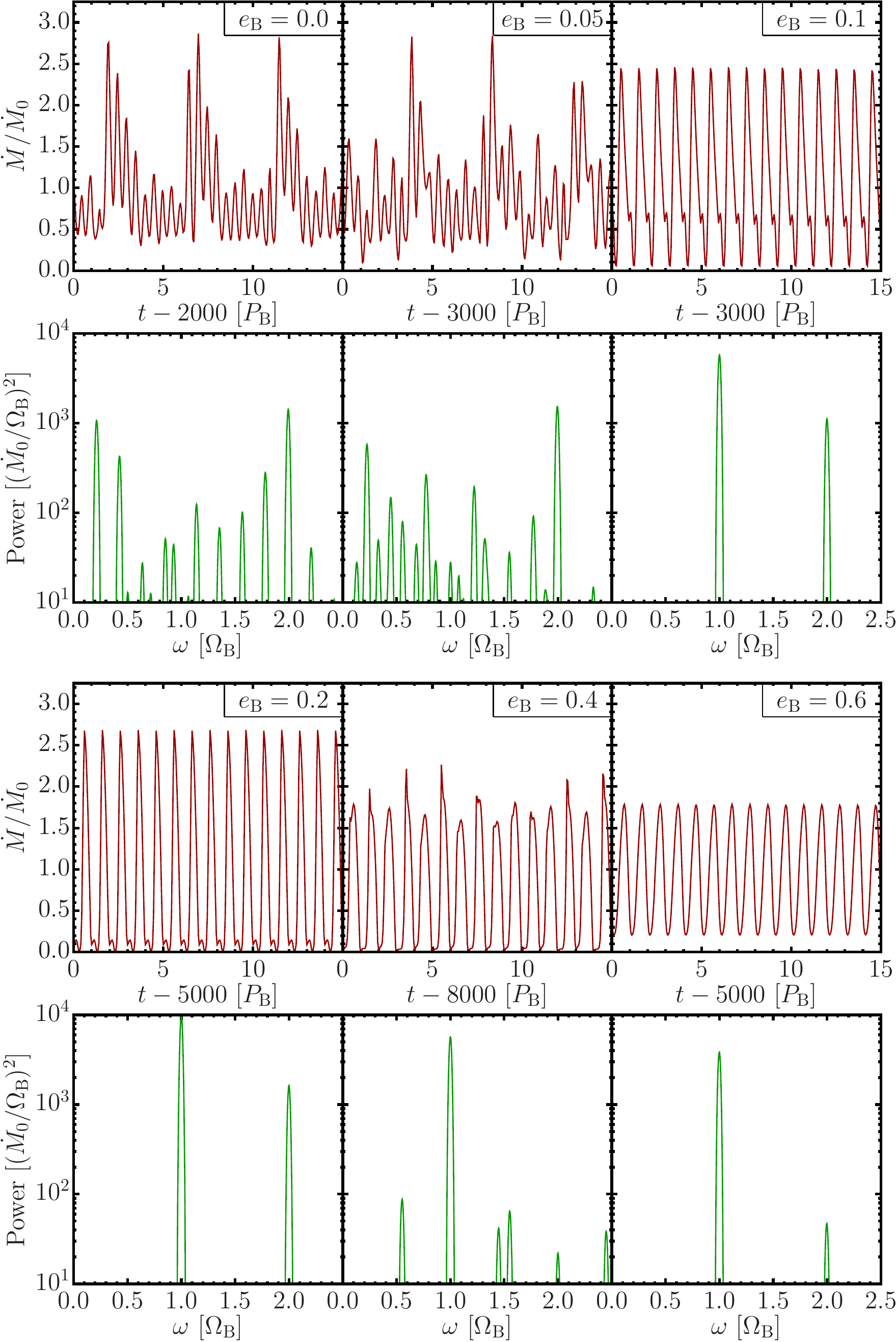}
\caption{The mass accretion rate at $r_\mathrm{in} = (1+e_\mathrm{B})a_\mathrm{B}$ for an equal mass binary ($q_\mathrm{B}= 1.0$) and $\alpha = 0.1$, for a range of binary eccentricities, $e_\mathrm{B}$ (first and third rows), and its power spectrum, taken over $50 P_\mathrm{B}$ (second and fourth rows). The top sequence of eccentricities demonstrates the transition from accretion modulated at a frequency of approximately $\Omega_\mathrm{B}/5$, to accretion modulated at $\Omega_\mathrm{B}$, which occurs at about $e_\mathrm{B} = 0.05$. The bottom sequence indicates the typical behavior for the moderate to high binary eccentricity cases, where the accretion is modulated only at $\Omega_\mathrm{B}$. See Section \ref{subsec:mdot_ecc}.}
\label{fig:mdot_power}
\end{center}
\end{figure}

\subsection{Dependence on Binary Eccentricity}
\label{subsec:mdot_ecc}

We explore the time-dependence of the local mass accretion rate\footnote{In this section, we use an approximation for $\dot{M}$, in which the (primitive) fluid variables $(\Sigma,\mathbf{u})$ are taken to be equal to their cell-averaged values at cell centres. These values, as output by \textsc{pluto}, are derived from the cell-averaged values of the conservative variables ($\Sigma$ and momentum density $\mathbf{m}$). In other words, the inter-cell variation of the primitive variables, which vanishes in the limit of infinite resolution, is ignored. This approximation is valid for exploring the short-term time-dependence of $\dot{M}$, because its variations are much larger than its time-averaged value. However, when evaluating small fluctuations in the time-averaged profile of $\dot{M}$ (as well as that of $\dot{J}$), as in Section \ref{sec:longterm_mdot_jdot}, the approximation breaks down. In this case, it is necessary to reconstruct the inter-cell variation, using the piecewise parabolic interpolation method (Colella \& Woodward 1984), as employed in the hydrodynamic solver, to evaluate the fluid variables at cell interfaces, before taking their appropriate products to calculate $\dot{M}$ or $\dot{J}$. Taking products of the cell-averaged fluid variables can result in anomalous features in the profiles of $\dot{M}$ and $\dot{J}$, wherever the variables have strong inter-cell gradients. The interpolation allows the different variables to be evaluated at the same location before their products are taken, resulting in more accurate profiles.},
\be
\dot{M}(r,t) = -\oint r \Sigma u_r \mathrm{d}\phi,
\ee
which is generally variable on orbital time-scales. The accretion rate as a function of $r$ and $t$ are shown in Fig.~\ref{fig:mdot_r_t}, for two representative cases ($e_\mathrm{B} = 0$ and $0.8$). For the circular case, two clear periodicities are evident. First, there are variations with a frequency of $2 \Omega_\mathrm{B}$ in the inner disc ($r \lesssim 4 a_\mathrm{B})$. Second, there are larger variations, with a frequency of about $\Omega_\mathrm{B}/5$, which are present out to about $\sim 10 a_\mathrm{B}$, having the largest amplitude at about $(2-3) a_\mathrm{B}$. The second type of variability has been seen in previous simulations of discs around circular binaries (e.g., MacFadyen \& Mirosavljevic 2008; Farris et al. 2014), and is sometimes referred to as having a frequency of $(2/9) \Omega_\mathrm{B}$ (note that we find the exact frequency to be $0.21 \Omega_\mathrm{B}$, with a FWHM in power of $0.03 \Omega_\mathrm{B}$, so the distinction between $1/5$ and $2/9$ is irrelevant). Note that in our simulations, $\dot{M}$ is always positive at the inner boundary, as required by the diode boundary condition. Mu\~{n}oz \& Lai (2016) showed that this feature is preserved in more realistic simulations that resolve the accretion streams onto individual bodies, at least for the $e_\mathrm{B} = 0$ case. Beyond about $10 a_\mathrm{B}$, $\dot{M} \approx \dot{M}_0$ at all times. From just beyond the inner boundary to about $10 a_\mathrm{B}$, the sign of $\dot{M}$ continuously alternates with tiime, indicating that, instantaneously, mass may be flowing inwards (towards the binary) or outwards (away from it). In a time-averaged sense, $\dot{M}$ is positive everywhere, so mass flows toward the binary on average (see Section \ref{sec:longterm_mdot_jdot}). For the $e_\mathrm{B} = 0.8$ case (see the bottom panel of Fig.~\ref{fig:mdot_r_t}), $\dot{M}$ varies only at $\Omega_\mathrm{B}$, and its maximum amplitude is about $5$ times smaller than for the circular case. As in the case of the circular binary, $\dot{M}$ is strictly positive near the inner boundary, as well as sufficiently far from the binary ($r \gtrsim 6 a_\mathrm{B}$), but can have either positive or negative sign at intermediate distances. This ``sloshing'' effect (fluctuation in the magnitude and sign of $\dot{M}$), which is largest at $\sim 3 a_\mathrm{B}$, and largest overall for circular binaries, is responsible for the variations in the azimuthally-averaged surface density profiles shown in Fig.~\ref{fig:density_profiles}.

Of particular interest is the variability of the mass accretion rate at $r_\mathrm{in}$, which we take to represent the actual accretion rate onto the binary (this is subject to the validity of the diode boundary condition). The different behaviors are illustrated in Fig.~\ref{fig:mdot_power}, which shows $\dot{M}(r_\mathrm{in},t)$, and its power spectrum, for equal mass binaries with various eccentricities. The top row demonstrates the transition from the circular binary behavior to the eccentric binary behavior. For $e_\mathrm{B} = 0$,  the same variabilities seen in Fig.~\ref{fig:mdot_r_t} are evident: large amplitude ``bursty'' fluctuations at about $\Omega_\mathrm{B}/5$, and smaller amplitude fluctuations at $2\Omega_\mathrm{B}$. For $e_\mathrm{B} = 0.05$, the behavior is similar to that of the $e_\mathrm{B} = 0$ case, although the $\Omega_\mathrm{B}/5$ spike in the power spectrum is somewhat less distinct, and additional low-frequency components are present. For $e_\mathrm{B} = 0.1$, the variability behavior becomes drastically different, and is dominated by oscillations at $\Omega_\mathrm{B}$ and $2\Omega_\mathrm{B}$. The power spectrum is much cleaner than for the lower values of $e_\mathrm{B}$, only having peaks at these two frequencies. Therefore, the accretion rate only exhibits low-frequency (sub-$\Omega_\mathrm{B}$) variability for $e_\mathrm{B} \lesssim 0.05$. The bottom two panels of Fig.~\ref{fig:mdot_power} demonstrate that the simple variability at only $\Omega_\mathrm{B}$ is the typical behavior for all higher values of $e_\mathrm{B}$. The exception is the case of $e_\mathrm{B} = 0.4$, for which the accretion rate variability appears less clean, and has small additional components  present in the power spectrum (at about $0.5$ and $1.5$ times $\Omega_\mathrm{B}$). This intermediate $e_\mathrm{B}$ is also associated with other unique behaviors, including periapse locking of the disc with the binary (see Section \ref{sec:eccentricity_precession}), large gravitational torques, and low level of angular momentum transfer to the binary (see Section \ref{sec:longterm_mdot_jdot}).

\subsection{Lump at Inner Edge of Disc}
\label{subsec:lump}

\begin{figure*}
\begin{center}
\includegraphics[width=0.99\textwidth,clip]{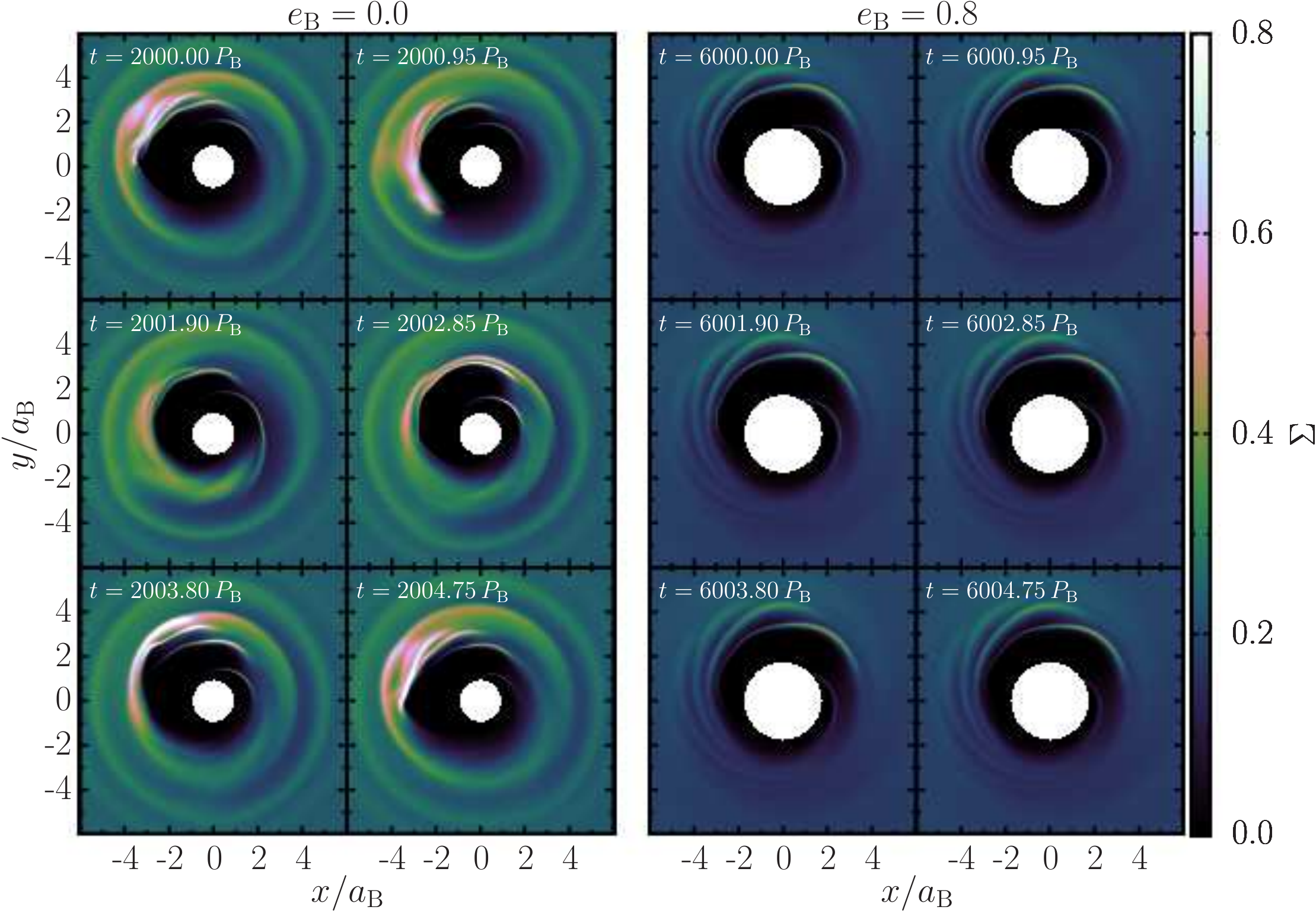}
\caption{\textit{Left}: Surface density snapshots illustrating one cycle of the creation and destruction of the lump at the inner edge of the disc for a circular binary. The six snapshots are evenly separated by $1/5$ of the approximate period of the lowest frequency component of the power spectrum of $\dot{M}(r_\mathrm{in})$ (the exact period is $4.72 P_\mathrm{B}$). In the first two panels, the lump is visible, orbiting at approximately the local Keplerian frequency. In the next panel, it has been sheared apart, and only a trace of it is left. This panel corresponds to the time at which the mass accretion rate onto the binary reaches its peak. In the fourth and fifth panels, the lump can be seen re-forming due to a pileup of streams emanating from the central cavity. Finally, in the last panel, the lump has reappeared, and is very close to its original position shown in the first panel. \textit{Right}: Same as the left panels except for an $e_\mathrm{B} = 0.8$ binary. The relative steadiness of the disc over the same interval of time can be clearly seen. Spiral density waves, which pile up at the disc apocentre (near the top of each panel), are also visible. They appear very similar in different snapshots, which are separated by approximately one binary orbital period. However, small differences of the phase of the innermost density wave/accretion stream can be seen because they are not exactly one orbit apart. No sign of a lump can be seen in this case. The presence or absence of a lump correlates with the presence or absence of $\Omega_\mathrm{B}/5$ modulation of the accretion rate. See Section \ref{subsec:lump}.}
\label{fig:lump}
\end{center}
\end{figure*}

\begin{figure*}
\begin{center}
\includegraphics[width=0.749\textwidth,clip]{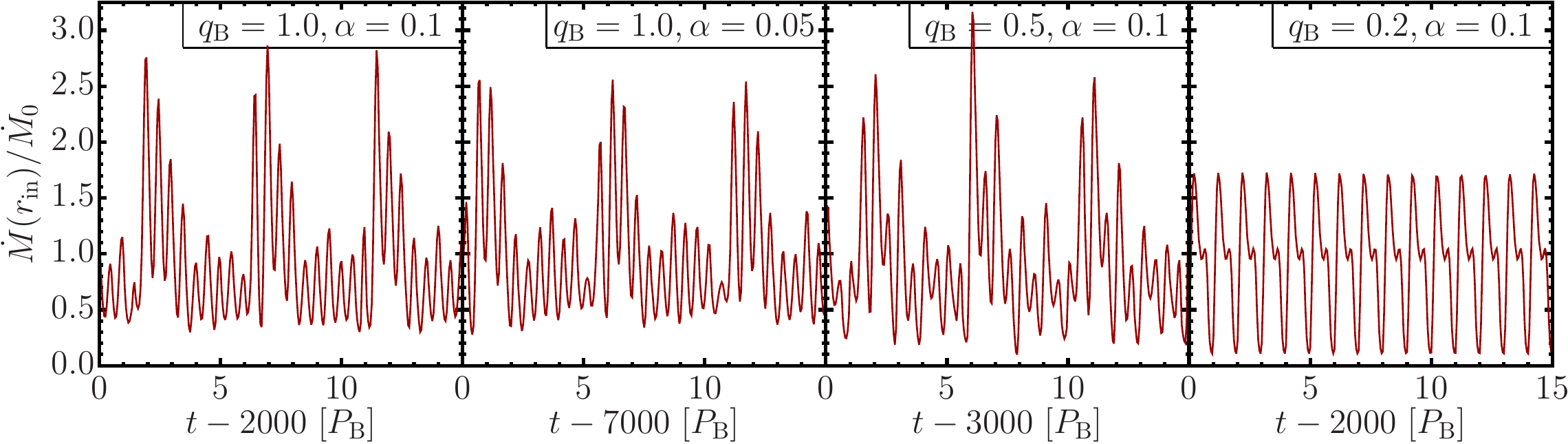}
\caption{Mass accretion rate as a function of time (as in Fig.~\ref{fig:mdot_power}, but without the associated power spectra) for circular binaries ($e_\mathrm{B} = 0$), for different mass ratios, and different values of $\alpha$. The leftmost panel, with $q_\mathrm{B}  = 1.0$ and $\alpha = 0.1$, is repeated from Fig.~\ref{fig:mdot_power}. For equal mass binaries, low-frequency ($\Omega_\mathrm{B}/5$) accretion modulation occurs indepedently of the value of $\alpha$. For non-equal mass binaries, it also occurs, although somewhat modified, for $q_\mathrm{B} = 0.5$, but is gone for $q_\mathrm{B} = 0.2$, and is replaced instead by modulation at $\Omega_\mathrm{B}$ only. See Section \ref{subsec:mdot_q}.}
\label{fig:mdot_2}
\end{center}
\end{figure*}

The variability of $\dot{M}(r_\mathrm{in},t)$ for circular and slightly eccentric binaries ($e_\mathrm{B} \lesssim 0.05$) is associated with the presence of a lump at the inner edge of the disc, which is not present for higher values of $e_\mathrm{B}$. The periodic creation and destruction of this lump, with a period corresponding to the Keplerian orbital period at about  $2.8 a_\mathrm{B}$, is temporally coincident with the modulation of $\dot{M}(r_\mathrm{in},t)$ with the same period (MacFadyen \& Milosavljevi{\'c} 2008; Shi et al. 2012; D'Orazio et al. 2013; Farris et al. 2014; Mu{\~n}oz \& Lai 2016). The lump cycle is illustrated in the left two panels of Fig.~\ref{fig:lump} (note that the time interval shown here corresponds to the first $5$ orbits shown in the leftmost panel of Fig.~\ref{fig:mdot_power}). After its creation, the lump rotates for about a third of an orbit before it is torn apart and flung towards the inner boundary. A new lump is then created by the accumulation of streams from the central cavity, and the cycle repeats. In Fig.~\ref{fig:snapshots}, a lump, though slightly less well-defined, can also be seen for $e_\mathrm{B} = 0.05$, which also displays low-frequency ($\sim \Omega_\mathrm{B}/5$) accretion variability. For comparison, the right two panels of Fig.~\ref{fig:lump} show the disc around an eccentric binary over the same interval of time as shown in the left two panels. The lump feature is clearly not present. Only spiral density waves and accretion streams, which look remarkably similar from one orbit to the next, can be seen. This demonstrates why the variability of the accretion rate at $r_\mathrm{in}$ for eccentric binaries is essentially a clean sinusoid whose period matches that of the binary.

The fact that a lump appears in simulations with $e_\mathrm{B} \lesssim 0.05$, but not for larger values of $e_\mathrm{B}$, indicates that it is a sensitive structure. Discs around circular binaries experience gravitational forcings which have pattern frequencies equal to the binary orbital frequency only. For eccentric binaries, there are many additional forcings, having both larger and smaller pattern frequencies, which become stronger with increasing $e_\mathrm{B}$. The fact that the lump disappears above a very small value of $e_\mathrm{B}$ may indicate that it is easily destroyed (or prevented from forming) by these additional forcings.

\subsection{Dependence on Binary Mass Ratio}
\label{subsec:mdot_q}

Figure~\ref{fig:mdot_2} shows $\dot{M}(r_\mathrm{in},t)$ for simulations with circular binaries, with different values of $q_\mathrm{B}$ and $\alpha$. They all exhibit low-frequency ($\sim \Omega_\mathrm{B}/5$) variabilities, except for the case of $q_\mathrm{B} = 0.2$, whose accretion rate varies only at a frequency of $\Omega_\mathrm{B}$, with weak modulation at $2 \Omega_\mathrm{B}$, somewhat resembling the behavior of an eccentric binary. Figure~\ref{fig:snapshots} shows that this is also the only disc around a circular binary in our runs that does not have a lump feature. Therefore, the presence of a lump and associated low-frequency accretion variability not only requires a small binary eccentricity, but also a sufficiently large mass ratio ($q_\mathrm{B} \gtrsim 0.2$). This dependence on mass ratio was also pointed out by Farris et al. (2014).

\section{Long Time-Scale Variability: Disc Eccentricity and Precession}
\label{sec:eccentricity_precession}

\begin{figure*}
\begin{center}
\includegraphics[width=0.749\textwidth,clip]{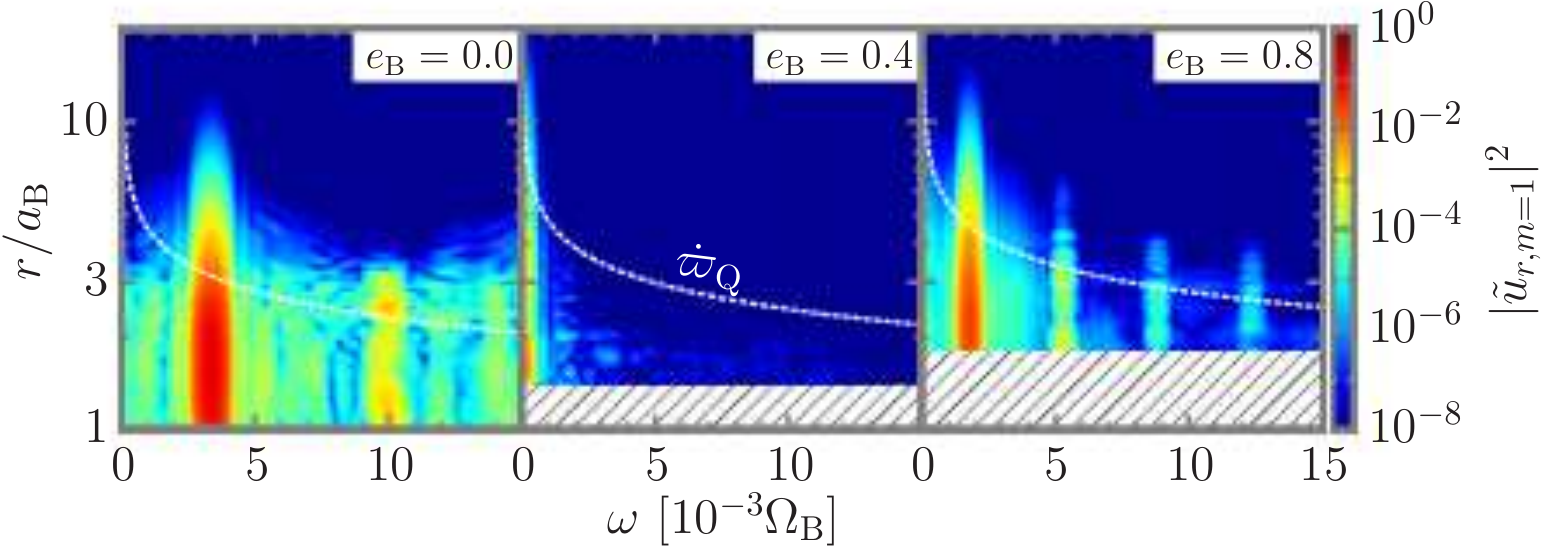}
\caption{Low-frequency power spectrum of the $m = 1$ Fourier component of $u_r$, taken over $2000$ binary orbits, for three different binary eccentricities (with $q_\mathrm{B} = 1.0$ and $\alpha = 0.1$ in all three cases). The dashed line in each panel corresponds to the precession frequency of a test particle due to the binary quadrupole potential, $\dot{\varpi}_\mathrm{Q}$ (Eq.~\ref{eq:omegadot_quad}). The presence of spatially coherent power for $e_\mathrm{B} = 0.0$ and $e_\mathrm{B} = 0.8$ corresponds to a global precessional mode. Such a mode is conspicuously absent for $e_\mathrm{B} = 0.4$ (note that the apparent power near zero frequency is spurious, as it corresponds to periods longer than the interval over which the power spectrum was taken). In the cases for which this mode is present, its frequency corresponds to $\dot{\varpi}_\mathrm{Q}$ at $r \approx (3 - 5) a_\mathrm{B}$, the approximate location of $r_\mathrm{peak}$.}
\label{fig:power_lowfreq}
\end{center}
\end{figure*}

\begin{figure*}
\begin{center}
\includegraphics[width=0.749\textwidth,clip]{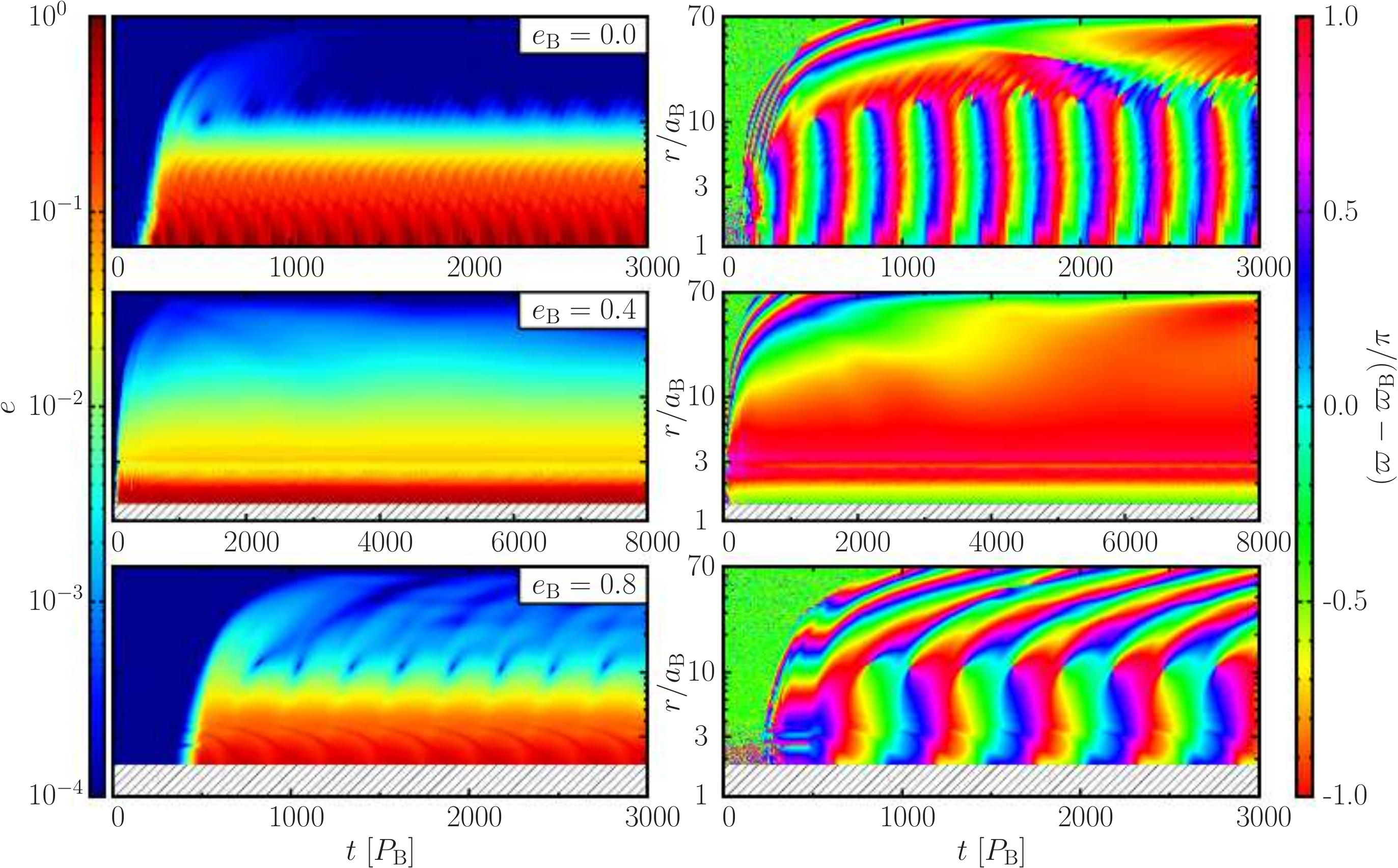}
\caption{Examples of the two different behaviors of the disc eccentricity (precession and apsidal locking) for different binary eccentricities (see Section \ref{sec:eccentricity_precession}). The disc eccentricity (magnitude of the mass weighted, azimuthally averaged eccentricity vector; left panels) and argument of pericentre (phase of the mass weighted, azimuthally averaged eccentricity vector; right panels) are shown as a function of time ($x$-axis) and radius ($y$-axis), for different values of $e_\mathrm{B}$ ($0$, $0.4$, and $0.8$; different rows), all for $q_\mathrm{B} = 1$ and $\alpha = 0.1$. The hatching in the bottom of the plots in the middle and bottom rows indicate that the computational domain does not extend all the way to $r/a_\mathrm{B} = 1$ (instead, the inner disc edge is at $r_\mathrm{in}/a_\mathrm{B} = 1+e_\mathrm{B}$). Coherent precession of the inner eccentric disc [$(2 - 10) a_\mathrm{B}$], as indicated by vertical strips in the right panels, is clearly seen for both $e_\mathrm{B} = 0$ and $0.8$. For $e_\mathrm{B} = 0.4$, the pericentre of the inner eccentric disc instead remains aligned with the binary (note that $\varpi - \varpi_\mathrm{B}$ equal to either $0$ or $\pm \pi$ indicates alignment, since the two values are degenerate for the equal mass binary shown here). Note that the range of the time axis is the same in the top and bottom panels (thus the precession periods can be compared visually), but it is more than twice as long in the middle panel, demonstrating that $\varpi$ is truly static in this case, on time-scales much longer than the precession periods seen in the other cases. The alignment for the $e_\mathrm{B} = 0.4$ case is also associated with smaller disc eccentricity (by a factor of $\sim 2$) at $r \sim (3 - 10) a_\mathrm{B}$.}
\label{fig:ecc_vector}
\end{center}
\end{figure*}

\begin{figure*}
\begin{center}
\includegraphics[width=0.749\textwidth,clip]{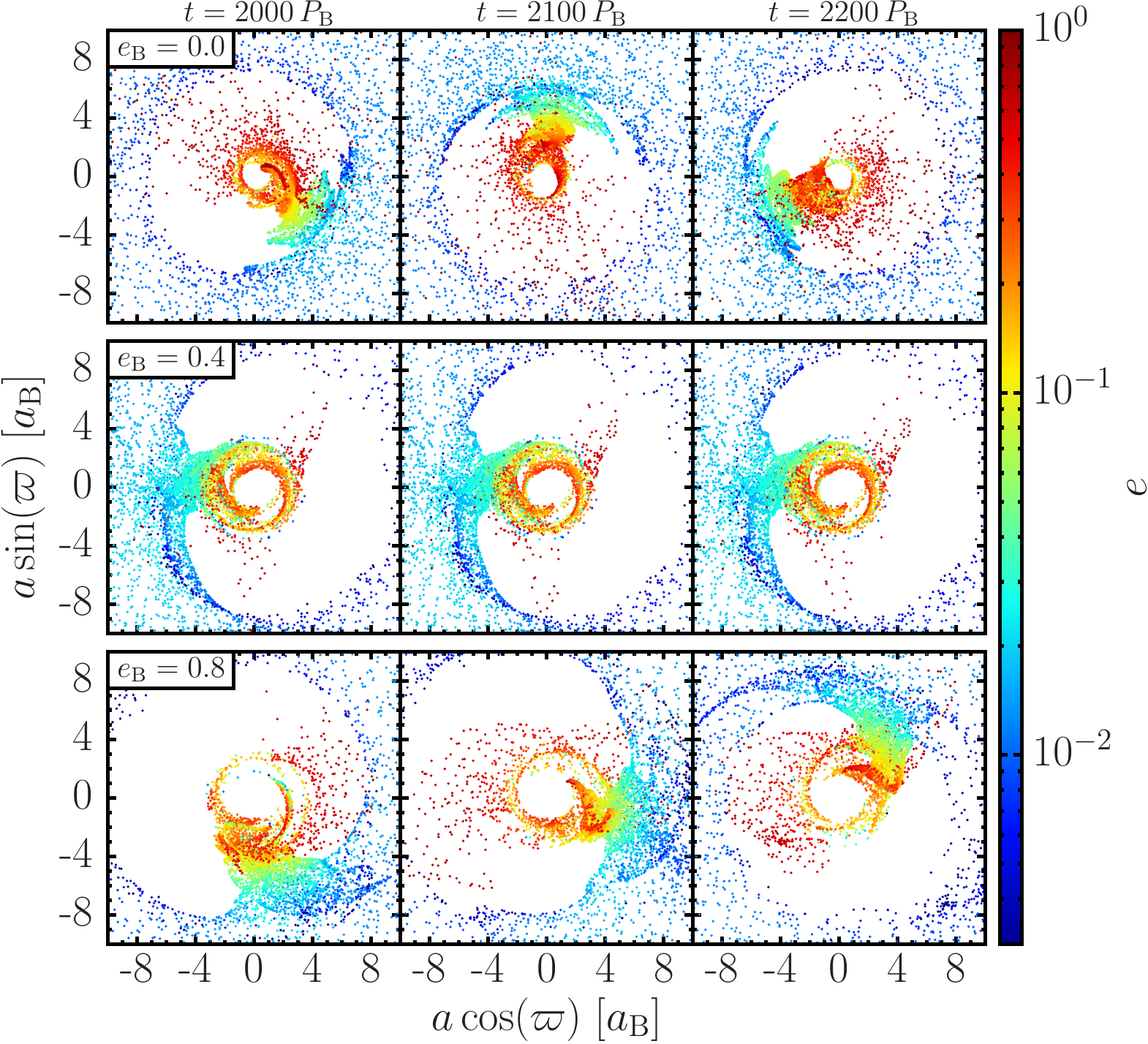}
\caption{Orbital elements of the fluid elements (grid cells) in the disc, only $5$ per cent of which are shown, at several points in time, for equal mass binaries with $e_\mathrm{B} = 0.0$ (top row), $0.4$ (middle row), and $0.8$ (bottom row). The radial distance of a point from the origin indicates its semi-major axis, its polar angle represents its pericentre orientation (where the binary line of apses is along the $x$-axis), and its colour indicates its eccentricity. In all three cases, there are three distinct populations of fluid elements: (i) those with nearly circular orbits ($e \lesssim 0.02$), large semi-major axes ($a \gtrsim 6 a_\mathrm{B}$), and no preferred pericentre orientation, corresponding to a circular disc; (ii) those with moderate eccentricites ($0.03 \lesssim e \lesssim 0.15$) and semi-major axes between $3 a_\mathrm{B}$ and $6 a_\mathrm{B}$, with pericentres clustered around one particular direction at a given time, representing a coherently eccentric part of the disc which precesses (top and bottom rows), or remains locked with the binary (middle row); and (iii) those with very high eccentricity ($e \approx 1$) and $a \lesssim 3 a_\mathrm{B}$, which represent material plunging toward the binary on nearly radial orbits (accretion streams). For $e_\mathrm{B} = 0.0$ and $0.8$, the streams are nearly isotropic in $\varpi$, while for $e_\mathrm{B} = 0.4$, they appear to have two preferred orientations that remain fixed in time.}
\label{fig:phase_space}
\end{center}
\end{figure*}

Having studied the short-term ($\sim P_\mathrm{B}$) variabilities of the circumbinary accretion in Section \ref{sec:short_timescale}, we now investigate variabilities of the disc on time-scales $\gg P_\mathrm{B}$. To this end, we examine the $m = 1$ Fourier component of the radial velocity $u_r$, and compute its power spectrum,
\be
|\tilde{u}_{r,m=1}|^2(r,\omega) = \left|\frac{1}{t_2-t_1}\int_{t_1}^{t_2} \oint u_r(r,\phi,t) \mathrm{e}^{\mathrm{i}(\phi -\omega t)} \mathrm{d}\phi \mathrm{d}t \right|^2,
\ee
where $t_1 = t_\mathrm{end} - 2000 P_\mathrm{B}$ to $t_2 = t_\mathrm{end}$. Figure~\ref{fig:power_lowfreq} shows the results for three different values of $e_\mathrm{B}$. We choose $u_{r,m=1}$ as a diagnosis of an eccentric disc, because an eccentric orbit is characterized by the $m = 1$ pattern in $u_r$, which may precess (rotate) at a frequency much slower than the local orbital frequency. The power spectrum of the same Fourier component of another fluid variable, such as $\Sigma$, would also reveal the same periodicity (due to mass continuity), but we choose $u_r$ since its equilibrium value is nearly zero (except for a small viscous drift). We see that for the $e_\mathrm{B} = 0.0$ and $e_\mathrm{B} = 0.8$ cases, the power spectrum is concentrated in a feature with a frequency of a few thousandths of $\Omega_\mathrm{B}$, which is coherent over a large range of radii. Weaker features at harmonics of this fundamental frequency can also be seen. Also shown in Fig.~\ref{fig:power_lowfreq} is the apsidal precession rate of a test particle (on a nearly circular orbit) due to the binary quadrupole potential (e.g., Liu et al. 2015a, Eq.~$20$),
\be
\label{eq:omegadot_quad}
\dot{\varpi}_\mathrm{Q} = \frac{3}{4}\frac{q_\mathrm{B}}{(1+q_\mathrm{B})^2}\left(1 + \frac{3}{2}e_\mathrm{B}^2\right)\left(\frac{r}{a_\mathrm{B}}\right)^{-7/2}\Omega_\mathrm{B}.
\ee
The features in the power spectrum for $e_\mathrm{B} = 0$ and $e_\mathrm{B} = 0.8$ have frequencies close to $\dot{\varpi}_\mathrm{Q}$ evaluated in the vicinity of $r_\mathrm{peak}$. Given the steep radial dependence ($r^{-7/2}$) of $\dot{\varpi}_\mathrm{Q}$, the exact value of $r$ at which the two quantities are equal is not particularly meaningful. However, the fact that the frequency of the feature is similar to the range of $\dot{\varpi}_\mathrm{Q}$ near the inner edge is significant, since it indicates that the feature is associated with the coherent precession of the inner disc. For $e_\mathrm{B} = 0.4$ (the middle panel of Fig.~\ref{fig:power_lowfreq}), this feature is conspicuously absent, indicating that the long-term dynamics of the disc are different in this case.

We investigate the eccentricity dynamics further by treating each grid cell as a test particle orbiting in the potential of the binary and, converting its instantaneous position $(r,\phi)$ and velocity $(u_r,u_\phi)$ into orbital elements $(a,e,\varpi)$, using energy conservation (the ``vis-viva equation''),
\be
a(r,\phi,t) = \left(\frac{2}{r} - \frac{u^2}{GM_\mathrm{B}}\right)^{-1},
\ee
and the (Runge-Lenz) eccentricity vector,
\be
\mathbf{e}(r,\phi,t) = \frac{\mathbf{u}^2\mathbf{r} - (\mathbf{u}\cdot\mathbf{r})\mathbf{u}}{GM_\mathrm{B}} - \hat{\mathbf{r}},
\ee
where $\mathbf{e} = [e\cos(\varpi),e\sin(\varpi)]$. We then use the mass-weighted azimuthal average of the eccentricity vector,
\be
\langle\mathbf{e}\rangle_\phi(r,t) = \frac{\oint \Sigma(r,\phi,t) \mathbf{e}(r,\phi,t)\mathrm{d}\phi}{\oint \Sigma(r,\phi,t)\mathrm{d}\phi},
\ee
to define the average eccentricity $e$ and pericentre angle $\varpi$ at each $r$, as a function of $t$. The result is shown in Fig.~\ref{fig:ecc_vector}, for simulations with three different binary eccentricities.

The eccentricity dynamics falls into one of two regimes. In the first regime ($e_\mathrm{B} = 0.0$ and $0.8$ in Fig.~\ref{fig:ecc_vector}), the eccentric portion of the disc undergoes coherent apsidal precession. The precession periods can be identified with those shown in Fig.~\ref{fig:power_lowfreq}. In the second regime ($e_\mathrm{B} = 0.4$), the eccentric disc keeps its line of apsides aligned with that of the binary. The distinction between the two regimes is unambiguous. In the second regime, not even a very slow pericentre advance, on the time-scale of the entire simulation ($8000$ orbits for the case shown in Fig.~\ref{fig:ecc_vector}), is seen. Instead, the disc pericentre stays truly aligned with that of the binary indefinitely. Apsidal alignment of the disc and binary was also reported by Lubow \& Artymowicz (2000) and Pierens \& Nelson (2007), although only for an unequal mass binaries. To our knowledge, apsidal alignment with an equal-mass binary--which does not have an octupole potential and cannot excite eccentricity in the secular regime (see Section \ref{subsec:test_particle})--has not been seen in numerical simulations before. Table \ref{tab:summary} indicates whether the disc precesses or remains aligned with the binary, as well as the precession period, where applicable, for each simulation.

The distinction between the precessing and aligned regimes is further highlighted in Fig.~\ref{fig:phase_space}, which shows snapshots of fluid elements in the disc in the $(a,e,\omega)$ phase space at several points in time, for the same cases as shown in Fig.~\ref{fig:ecc_vector}. In all three cases, the presence of a coherently eccentric inner disc manifests as the clustering of fluid elements in $\varpi$. In the precessing case, the cluster can be seen to rotate (by about $2/3$ and $1/3$ of a full rotation, for $e_\mathrm{B} = 0.0$ and $0.8$), over the $200 P_\mathrm{B}$ interval shown, while in the aligned case, the cluster remains at the same $\varpi$ as time progresses. In this figure, there is a distinction between two different populations of fluid elements, one with $e \sim 0.01 - 0.2$ (what we call the eccentric disc proper, between about $3$ and $6 a_\mathrm{B}$, represented by yellow/green dots), and another with $e \sim 1$ (red dots), which corresponds to streams penetrating the inner cavity on nearly radial orbits. In the precessing case, the streams do not have of a preferred orientation in $\varpi$, while for the aligned case, there are two strongly preferred directions (both $\sim 45^\circ$ from the binary line of apsides). This explains the apparent discontinuity in the average $\varpi$ of the disc close to the binary ($r/a_\mathrm{B} \lesssim 2$) seen in Fig.~\ref{fig:ecc_vector}: it is an artefact of the accretion streams, rather than of the eccentric disc proper, which does not exist close to the binary. The restricted orientation of the streams in the aligned case ($e_\mathrm{B} = 0.4$) may be associated with the reduced rate at which angular momentum is transferred to the binary (see Section \ref{sec:longterm_mdot_jdot}).

\section{Theoretical Explanations for Disc Eccentricity Excitation, Precession, and Apsidal Locking}
\label{sec:eccentricity_precession_theory}

\begin{figure*}
\begin{center}
\includegraphics[width=0.749\textwidth,clip]{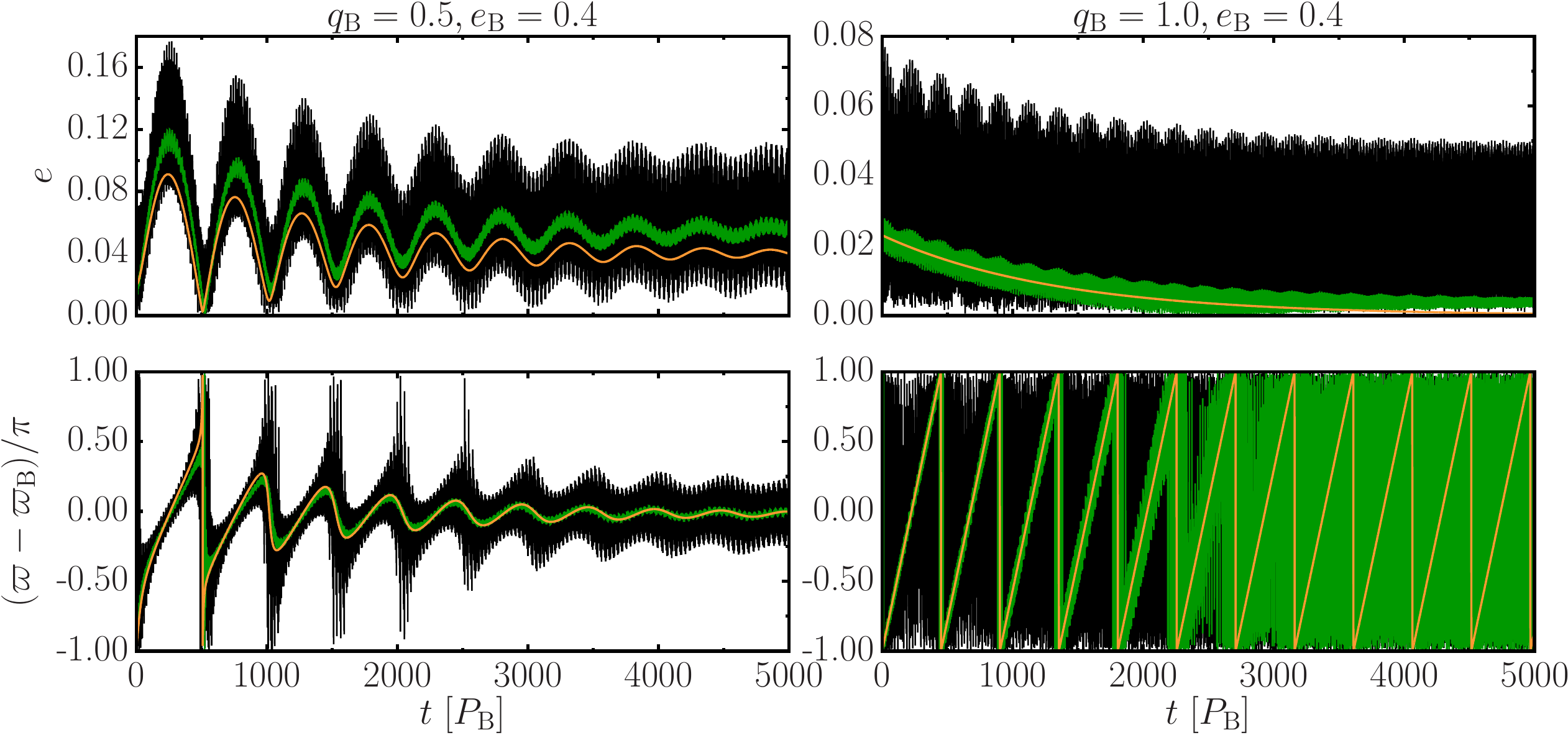}
\caption{Eccentricity (top) and pericentre longitude (bottom) evolution for a test particle with semi-major axis $a = 3.5 a_\mathrm{B}$ orbiting an eccentric binary with $e_\mathrm{B} = 0.4$ (see Section \ref{subsec:test_particle}). Two binary mass ratios are shown, $q_\mathrm{B} = 0.5$ (left), and $1.0$ (right). The orange lines are the results of integrations of the secular equations of motion (Eqs. \ref{eq:edot_secular}, \ref{eq:omegadot_secular}, and \ref{eq:edamp_secular}). The black lines are the results of integrations of the non-secular equation of motion (Eqs. \ref{eq:eom_non_secular} and \ref{eq:edamp_non_secular}), and the green lines indicate the $e$ and $\varpi$ of the running time average of the eccentricity vector (from the same integrations), taken over the interval $(t - 5P_\mathrm{B}, t + 5 P_\mathrm{B})$ for each $t$. The running average filters out short term variation in order to assess whether $\varpi$ is circulating or librating. For $q_\mathrm{B} = 0.5$, the balance of the eccentricity excitation by the binary octupole potential and viscous damping (modeled by Eq.~\ref{eq:edamp_secular} or Eq.~\ref{eq:edamp_non_secular}) results in the test particle settling into a finite eccentricity orbit with its pericentre aligned with that of the binary. For $q_\mathrm{B} = 1.0$, the lack of eccentricity excitation by an octupole potential results in pure damping of the eccenticity: the test particle evolves towards a circular orbit, while $\varpi$ continues to circulate indefinitely (although becoming increasingly less well-defined due to the very small $e$). The same behavior is seen for both the secular equations and non-secular equations, indicating that short time-scale dynamical forcings from the binary do not modify the long-term evolution of the particle.}
\label{fig:test_particle}
\end{center}
\end{figure*}

\begin{figure*}
\begin{center}
\includegraphics[width=0.749\textwidth,clip]{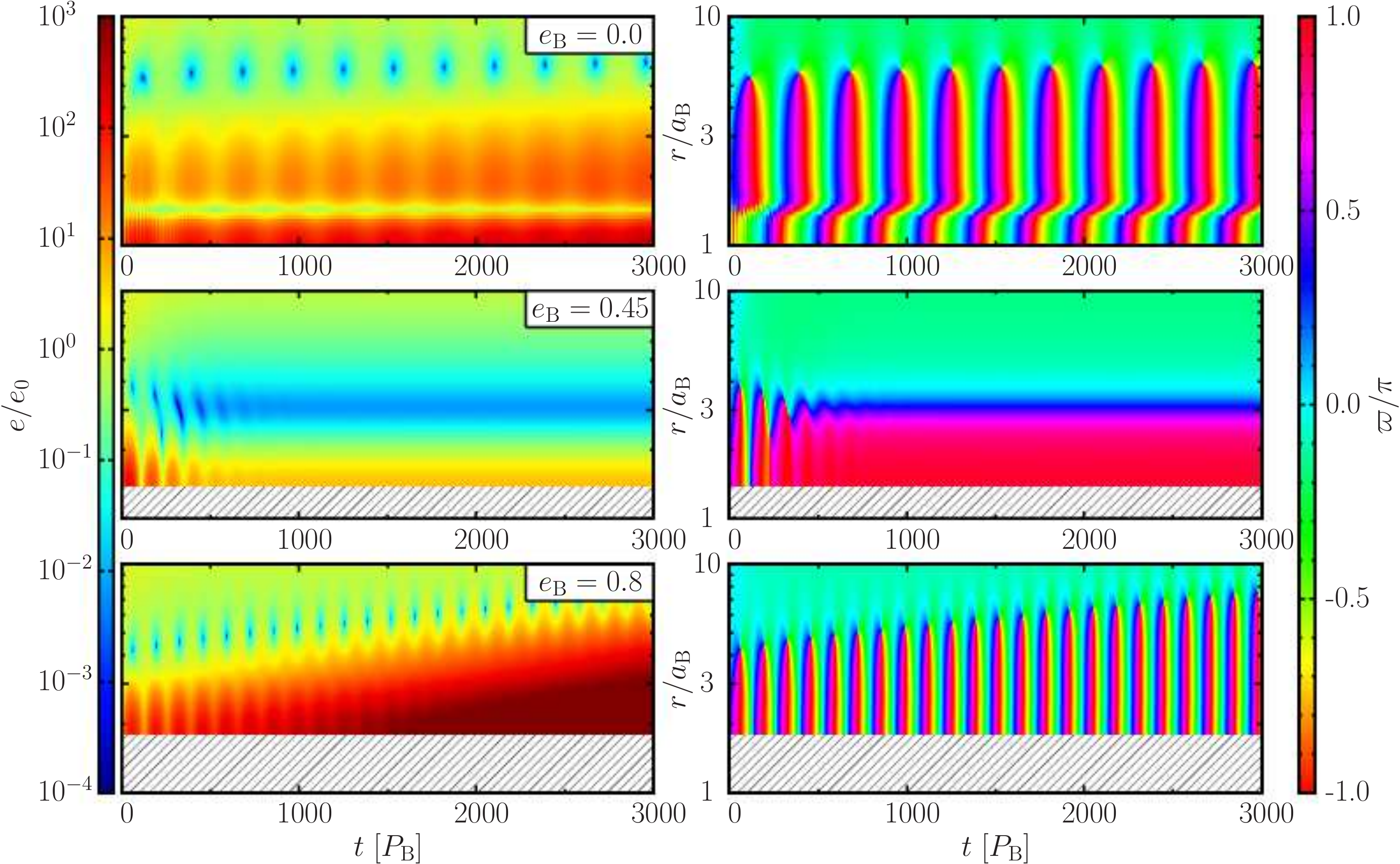}
\caption{Disc eccentricity and longitude of pericentre, as a function of time and radius (as in Fig.~\ref{fig:ecc_vector}), as calculated from the one dimensional linear fluid eccentricity equation (Eq.~\ref{eq:ecc_1d}), with eccentricity driven at two eccentric Lindblad resonances (ELRs; see Section \ref{subsec:1d_fluid}). Three different binary eccentricities are shown. In all cases, the binary has $q_\mathrm{B} = 1.0$, and the disc has $h = 0.1$ and $\alpha_\mathrm{b} = 0.1$ (as in our numerical simulations, except that the shear viscosity parameter $\alpha$ is replaced by a bulk viscosity parameter $\alpha_\mathrm{b}$). For $e_\mathrm{B} = 0.0$ (top), the disc eccentricity grows as a result of driving at the $\Omega=\Omega_\mathrm{B}/2$ ELR, and the disc precesses. Similarly, for $e_\mathrm{B} = 0.8$ (bottom), eccentricity grows as a result of driving at the $1/4$ ELR, and the inner disc [$(1.5 - 6) a_\mathrm{B}$] precesses coherently. For $e_\mathrm{B} = 0.45$, the eccentricity drivings at both ELRs are too weak to overcome viscous damping, so no eccentricity growth is observed, and the propagation of the eccentricity wave is damped, resulting in a halting of apsidal precession. This behavior may explain the similar qualitative trend seen in our numerical simulations, shown in Fig.~\ref{fig:ecc_vector}.}
\label{fig:ecc_1p1d}
\end{center}
\end{figure*}

In Section \ref{sec:eccentricity_precession}, we presented numerical results demonstrating that the inner region of a circumbinary disc generally becomes eccentric and evolves coherently. We showed that, for both low and high values of $e_\mathrm{B}$, the pericentre of the disc precesses coherently around the binary. However, for intermediate values of $e_\mathrm{B}$, the disc instead becomes apsidally locked with the binary. This result is puzzling and unexpected, and so in this section, we explore several possible theoretical explanations for this behavior. This section is self-contained, and can be skipped if the reader wishes to continue on to the presentation of the rest of our numerical results, which resumes in Section \ref{sec:longterm_mdot_jdot}.

\subsection{Test Particle Dynamics}
\label{subsec:test_particle}

We first take the simplest approach, considering the dynamics of a test particle which orbits the binary, and which is subject to the tidal potential of the binary as well as a parametrized ``frictional'' eccentricity damping force.

\subsubsection{Secular Dynamics}
The long-term dynamics of the particle may be described using the secular approximation, in which the orbital motion of both the binary and the particle are averaged out. The resulting orbital evolution of a free particle, to octupole order, is given by (e.g., Liu et al. 2015b\footnote{We have corrected typos in the relevant equations of Liu et al. (2015b): in Eq.~($12$), $(1-e_2^2)^{5/2}$ should be $(1-e_2^2)^2$, and in Eq.~($14$), $(4 + 3e_1)$ should be $(4 + 3e_1^2)$.})
\be
\label{eq:edot_secular}
\frac{1}{\Omega_\mathrm{K}}\frac{\mathrm{d}e}{\mathrm{d}t} = -\frac{15}{64}\frac{q_\mathrm{B}(1-q_\mathrm{B})}{(1+q_\mathrm{B})^3}\left(\frac{a_\mathrm{B}}{a}\right)^3\frac{e_\mathrm{B}(4+3e_\mathrm{B}^2)}{(1-e^2)^2}\sin(\varpi-\varpi_\mathrm{B}),
\ee
\be
\label{eq:omegadot_secular}
\begin{aligned}
\frac{1}{\Omega_\mathrm{K}}\frac{\mathrm{d}\varpi}{\mathrm{d}t} &= \frac{3}{8}\frac{q_\mathrm{B}}{(1+q_\mathrm{B})^2}\left(\frac{a_\mathrm{B}}{a}\right)^2\frac{(2+3e_\mathrm{B}^2)}{(1-e^2)^2} \\ 
&- \frac{15}{64}\frac{q_\mathrm{B}(1-q_\mathrm{B})}{(1+q_\mathrm{B})^3}\left(\frac{a_\mathrm{B}}{a}\right)^3\frac{e_\mathrm{B}(4+3e_\mathrm{B}^2)(1+4e^2)}{e(1-e^2)^3} \\
& \times \cos(\varpi-\varpi_\mathrm{B}).
\end{aligned}
\ee
In general, the particle and the binary exchange angular momentum (although the amount gained or lost by the binary is negligible in the test particle limit), resulting in oscillations of the eccentricity of the particle, which in the linear regime ($e,e_\mathrm{B} \ll 1$), has an amplitude of
\be
e_\mathrm{forced} = \frac{5}{8}\left(\frac{1-q_\mathrm{B}}{1+q_\mathrm{B}}\right)\frac{a_\mathrm{B}}{a}\left(\frac{4+3e_\mathrm{B}^2}{2+3e_\mathrm{B}^2}\right)e_\mathrm{B}
\ee
(e.g., Moriwaki \& Nakagawa 2004). To model the effect of eccentricity damping (e.g., due to viscosity), we add the following term to the right hand side of Eq.~(\ref{eq:edot_secular}), 
\be
\label{eq:edamp_secular}
\dot{e}_\mathrm{damp} = -\frac{e}{t_\mathrm{d,s}},
\ee
where $t_\mathrm{d,s}$ is the eccentricity damping time-scale. In the presence of eccentricity damping, the particle can evolve towards a fixed state in which $\varpi - \varpi_\mathrm{B} = 0$ and $e \approx e_\mathrm{forced}$ (e.g., Wu \& Goldreich 2002). This apsidal alignment can occur only if $e_\mathrm{forced} \neq 0$, which requires that $q_\mathrm{B} < 1$ and $e_\mathrm{B} > 0$ (i.e., the binary octupole potential is non-zero), otherwise the eccentricity of the test particle will continually precess as its eccentricity asymptotically approaches zero.

Figure~\ref{fig:test_particle} shows several examples of the secular evolution of a test particle subject to eccentricity damping. In these examples, the semi-major axis of the particle is chosen to be $a = 3.5 a_\mathrm{B}$, and the eccentricity damping time is $t_\mathrm{d,s} = 15/\dot{\varpi}_\mathrm{Q}$ (see Eq.~\ref{eq:omegadot_quad}). For the unequal mass binary, the argument of pericentre of the test particle undergoes several precession cycles, before librating around that of the binary with a decreasing amplitude, until it is essentially aligned with the binary. The eccentricity of the particle oscillates and eventually approaches a fixed value of $0.04 \approx e_\mathrm{forced}$. For the equal-mass binary, the eccentricity simply approaches zero, while the argument of pericentre precesses indefinitely.

In this secular framework, an unequal mass binary is required to produce apsidal alignment. This may explain the alignment seen in one of our numerical simulations, with $(q_\mathrm{B}, e_\mathrm{B}, \alpha) = (0.2, 0.4, 0.1)$. However, it cannot explain the alignment seen in several of our simulations with equal mass binaries.

\subsubsection{Non-Secular Dynamics}
Since secular theory cannot produce apsidal alignment without a non-zero octupole potential, we now consider the non-secular (i.e., non-orbit-averaged) dynamics of a test particle. This is potentially useful, since the fluid elements of the eccentric disc have semi-major axes which are only a few times that of the binary ($a/a_\mathrm{B} \approx 3 - 6$), so the short-term (orbital time-scale) forcings from the binary are not necessarily negligible compared to the long-term, secular forcings. The short-term forcings may give rise to additional eccentricity excitation which is not captured in the secular approximation.

The equation of motion of a test particle (with position vector $\mathbf{r}$) reads
\be
\label{eq:eom_non_secular}
\ddot{\mathbf{r}} = -\mathbf{\nabla} \Phi + \mathbf{f}_\mathrm{d},
\ee
where $\Phi$ is the (time-dependent) gravitational potential of the central binary. We take the damping force $\mathbf{f}_\mathrm{d}$ to be of the form
\be
\label{eq:edamp_non_secular}
\mathbf{f}_\mathrm{d} = -\frac{1}{t_\mathrm{d,ns}}\left[3(\hat{\mathbf{r}} \cdot \dot{\mathbf{r}})\hat{\mathbf{r}} + (\hat{\mathbf{r}} \times \dot{\mathbf{r}} - r\mathbf{\Omega}_\mathrm{p}) \times \hat{\mathbf{r}}\right],
\ee
with $\mathbf{\Omega}_\mathrm{p} = (|\dot{\mathbf{r}}|/|\mathbf{r}|)\hat{\mathbf{z}}$ (e.g., Mardling \& Lin 2002). This form of $\mathbf{f}_\mathrm{d}$ ensures that a circular orbit does not decay. We find that choosing $t_\mathrm{d,ns} = 2.5 t_\mathrm{d,s}$ results in the same eccentricity damping rates between the secular and non-secular equations. Thus we adopt $t_\mathrm{d,ns} = 37.5/\dot{\varpi}_\mathrm{Q}$ in the following numerical examples.

Figure~\ref{fig:test_particle} shows some numerical results based on integrating Eq.~(\ref{eq:eom_non_secular}). The particle starts at $3.5 a_\mathrm{B}$, with a small eccentricity. The eccentricity and argument of pericentre exhibit large amplitude, short-time-scale fluctuations, but on average are very similar to the corresponding secular results. In particular, for a non-equal mass binary, the particle becomes apsidally aligned with the binary, with a finite eccentricity, although still exhibiting fast, order unity fluctuations. The average value of the eccentricity is slightly larger than in the secular calculation due to these fluctuations. For the equal mass binary, the orbital evolution of the test particle is also very similar to the secular calculation. Despite the large amplitude, fast fluctuations, the initial eccentricity of the particle is smoothly damped, and its argument of pericentre continues to precess. Although the average eccentricity becomes vanishingly small, it can still be as large as $0.05$ instantaneously, indicating that there is some some additional eccentricity excitation by the non-secular forcings from the binary. However, there is no sign of apsidal alignment for the equal mass binary. We conclude that the additional short-term forcings cannot explain the apsidal alignment of a disc around an equal mass binary.

We note that the results presented in Fig.~\ref{fig:test_particle} do not include the effect of mean motion resonances, which likely plays an important role in eccentricity excitation of the disc. In the following subsection, we consider the dynamics of a fluid disc, including resonant eccentricity excitation.

\subsection{Linear Fluid Dynamics of Eccentric Discs}
\label{subsec:1d_fluid}

The dynamics of an eccentric fluid disc can be formally described using linear perturbation theory, in which eccentricity propagates as a slow, one-armed ($m = 1$) density wave. The equation describing the evolution of the complex eccentricity vector $E = e\exp(\mathrm{i}\varpi)$ is (Goodchild \& Ogilvie 2006)
\be
\begin{aligned}
\label{eq:ecc_1d}
2r\Omega\frac{\partial E}{\partial t} & = -\frac{\mathrm{i}E}{r}\frac{\partial}{\partial r}\left(r^2\frac{\partial \Phi_2}{\partial r}\right) + \frac{\mathrm{i}E}{\Sigma}\frac{\partial P}{\partial r} \\
& + \frac{\mathrm{i}}{r^2\Sigma} \frac{\partial}{\partial r}\left[(1-\mathrm{i}\alpha_\mathrm{b})Pr^3\frac{\partial E}{\partial r}\right] \\
& + \sum_i 2a_\mathrm{B}\gamma_i r\Omega E \delta(r-r_{\mathrm{res},i}),
\end{aligned}
\ee
where
\be
\Phi_2 = -\frac{GM_\mathrm{B}}{4a_\mathrm{B}}\frac{q_\mathrm{B}}{(1+q_\mathrm{B})^2}\left(1+\frac{3}{2}e_\mathrm{B}^2\right)\left(\frac{r}{a_\mathrm{B}}\right)^{-3}
\ee
is the (time-independent) quadrupole component of the binary potential. The first two terms on the right-hand side of Eq.~(\ref{eq:ecc_1d}) describe precession due to the potential and gas pressure, respectively, and the third term describes the diffusion of eccentricity through the disc (with $\alpha_\mathrm{b}$ characterizing the disc viscosity). The fourth term describes the growth of eccentricity at various resonances, each having a growth rate $\gamma_i$, and which are idealized as infinitely narrow (i.e., $\delta$ functions) in $r$.

The binary potential can be decomposed into Fourier components,
\be
\Phi(r,\phi,t) = \sum_{m,N} \Phi_{m,N}\cos(m\phi-N\Omega_\mathrm{B}t),
\ee
where $\Phi_{m,N}$ is the strength of the potential component having azimuthal number $m$, rotating with pattern frequency $\omega_\mathrm{P} = N\Omega_\mathrm{B}/m$. Eccentricity can be excited in the disc at eccentric Lindblad resonances (ELRs), whose locations are determined by the criterion
\be
\Omega = \frac{m\omega_\mathrm{P}}{m+2},
\ee
and which have growth rates $\gamma_{m,N}$ that are proportional to $\Phi_{m,N}^2$ (Lubow 1991). We specialize to equal-mass binaries, which have a vanishing $m = 1$ potential component, and for which the $1$:$3$ ELR (e.g., Lubow 1991; Papaloizou et al.~2001; Fleming \& Quinn 2017) does not exist. Thus, resonances with $m = 2$ are the strongest, since $\Phi_{m,N} \propto r^{-m-1}$. For a circular binary, only the $\Phi_{m,m}$'s are non-zero, i.e., the only possible pattern frequency is $\omega_\mathrm{P} = \Omega_\mathrm{B}$. Thus, the most important resonance corresponds to the $\Omega = \Omega_\mathrm{B}/2$ commensurability (located at $r = 1.59 a_\mathrm{B}$), and its growth rate is proportional to $\Phi_{2,2}^2$. Resonances with higher values of $m$ are not important, as they are located very close to the binary, where the flow deviates strongly from Keplerian. For eccentric binaries, resonances with different pattern frequencies exist. Those with $\omega_\mathrm{P} > \Omega_\mathrm{B}$ are not relevant, since they are located inside the disc truncation radius. However, those with $\omega_\mathrm{P} < \Omega_\mathrm{B}$ are located at larger radii, and can be important. The second most important resonance is associated with $\omega_\mathrm{P} = \Omega_\mathrm{B}/2$, which is located at the $\Omega = \Omega_\mathrm{B}/4$ commensurability ($r = 2.52 a_\mathrm{B}$) and has a growth rate proportional $\Phi_{2,1}^2$.

The strengths of the $m = 2$ potential components, to quadrupole order, evaluated at the ELRs, are
\be
\Phi_{2,N}\left(r_\mathrm{ELR}\right) \approx -\frac{3N^2}{64} C_N \frac{GM_\mathrm{B}}{a_\mathrm{B}} \frac{q_\mathrm{B}}{(1+q_\mathrm{B})^2},
\ee
where the relevant coefficients, to order $e_\mathrm{B}^2$, are given by $C_2 \approx 1 - 5e_\mathrm{B}^2/2$ and $C_1 \approx -3e_\mathrm{B}$ (note that $C_N$ corresponds to the notation $C_{2,2,N-2}^\mathrm{CB}$ used in Miranda \& Lai 2015). For simplicity, we choose
\be
\gamma_0 \equiv \left. \gamma_{2,2}\right|_{e_\mathrm{B} = 0} = 2\frac{q_\mathrm{B}^2}{(1+q_\mathrm{B})^4}\Omega_\mathrm{B}
\ee
as an ansatz for the normalization of the ELR growth rates, using the result of Lubow (1991), where we have set the factors related to the disc geometry and resonance width to unity. We then have
\be
\gamma_{2,2} = \left(1-5e_\mathrm{B}^2\right)\gamma_0,
\ee
for $e_\mathrm{B} < (1/5)^{1/2}$ (otherwise $\gamma_{2,2} = 0$), and similarly,
\be
\gamma_{2,1} = \frac{9}{16}e_\mathrm{B}^2\gamma_0.
\ee

We investigate the dynamics of eccentric discs driven by these two ELRs by integrating Eq.~(\ref{eq:ecc_1d}), discretized on a uniformly spaced grid (with $\Delta r = 0.05 a_\mathrm{B}$), using a fourth-order Runge-Kutta method. As in our simulations, we adopt the locally isothermal equation of state, $c_\mathrm{s}(r) = 0.1 r\Omega_\mathrm{K}$, viscosity parameter $\alpha_\mathrm{b} = 0.1$, and choose the inner boundary to be $r_\mathrm{in} = (1+e_\mathrm{B})a_\mathrm{B}$. The outer boundary is at $10 a_\mathrm{B}$. Zero gradient conditions are imposed in the complex eccentricity ($\partial E/\partial r = 0$) at both boundaries, and a damping zone, which relaxes $E$ to its initial condition on the orbital time-scale of the outer boundary, is imposed in the outer $25$ per cent of the domain in order to mimic an outgoing wave boundary condition. The background surface density is adapted from the time averaged, azimuthally averaged profiles shown in Fig.~\ref{fig:density_profiles}, for the appropriate binary parameters.

The results of our one dimensional, eccentric fluid disc experiments are shown in Fig.~\ref{fig:ecc_1p1d}. For $e_\mathrm{B} = 0$, the disc eccentricity grows with time, and its pericentre precesses coherently. The precession period is similar to the one seen in the numerical simulations. A kink feature in $\varpi$ at the location of the inner resonance, similar to the one seen in the full numerical simulations (Fig.~\ref{fig:ecc_vector}), is also reproduced. For $e_\mathrm{B} = 0.45$ (this value is chosen because we found that the disc still precesses for $e_\mathrm{B} = 0.4$ in our 1D model), the eccentricity does not grow, but rather the initial profile of $e$ is reconfigured into a new equilibrium profile, which does not subsequently change with time. The argument of pericentre of the inner disc can be seen to precess through several cycles, with a shorter period than for $e_\mathrm{B} = 0$, before eventually settling into a fixed (but $r$-dependent) orientation. For $e_\mathrm{B} = 0.8$, the disc eccentricity again grows, and the disc precesses, although with a period that is several times shorter than the one seen in the numerical simulations. Thus, our 1D eccentric disc model based on Eq.~(\ref{eq:ecc_1d}) with two resonant driving terms captures many of the key features seen in the full numerical simulations. 

The behavior of our 1D model can be qualitatively understood as follows. In the local (WKB) limit, $E \propto \mathrm{e}^{\mathrm{i}\omega t - \mathrm{i}kr}$, far from the central binary, and far from any resonances, Eq.~(\ref{eq:ecc_1d}) reduces to
\be
\omega = -\frac{(1-\mathrm{i}\alpha_\mathrm{b})}{2}\frac{k^2 c_\mathrm{s}^2}{\Omega},
\ee
which is the dispersion relation for local spiral density waves, $(\omega-m\Omega)^2 = \Omega^2 + k^2c_\mathrm{s}^2$, in the limit of low frequency ($\omega \ll \Omega$), and with $m = 1$, with the addition of viscous damping. If the strength of eccentricity driving at a resonance is not strong enough to overcome viscous damping, then not only will the eccentricity not grow, but the wave will not propagate, meaning that no precession will occur. The strength of the the inner ($\Omega/\Omega_\mathrm{B} = 1/2$) resonance is a decreasing function of $e_\mathrm{B}$, while that of the outer resonance ($\Omega/\Omega_\mathrm{B} = 1/4$) is an increasing function of $e_\mathrm{B}$. Therefore, for small values of $e_\mathrm{B}$ (including zero), the inner resonance is strong and can effectively drive eccentricity growth and precession in the disc. For sufficiently large values of $e_\mathrm{B}$, the inner resonance is weak (or has shut off entirely), while the outer resonance is strong, and can drive eccentricity growth and precession. For intermediate $e_\mathrm{B}$, neither resonance is strong, and so the complex eccentricity $E$ freezes out. Thus, our 1D model qualitatively explains two features in our numerical simulations: i) the lack of precession for intermediate values of $e_\mathrm{B}$, and ii) the associated reduced disc eccentricity, compared to the cases of low or high values of $e_\mathrm{B}$.

There are several caveats to consider when comparing our 1D eccentric disc models (Fig.~\ref{fig:ecc_1p1d}) and full 2D simulations (Fig.~\ref{fig:ecc_vector}). First, the linear treatment in the 1D model only captures eccentricity growth (which occurs for precessing discs) or non-growth (for non-precessing discs), and cannot capture its saturation, which presumably occurs due to non-linear effects. In our 2D simulations, eccentricity always grows and settles to some equilibrium value (actually a range of values at different locations in the disc), although it is smaller by approximately a factor of $2$ in the apsidally aligned regime, compared to in the precessing regime. Second, the argument of pericentre $\varpi$ in Eq.~(\ref{eq:ecc_1d}) is only defined relative to an arbitrary reference angle, rather than with respect to the binary argument of pericentre $\varpi_\mathrm{B}$. This is a consequence of considering the binary gravitational potential only to quadrupole order (this is appopriate, since here we are focusing on equal mass binaries, for which the octupole potential vanishes). As a result, when precession of the disc is halted in our 1D model (the middle panel of Fig.~\ref{fig:ecc_1p1d}), the value of $\varpi$ at which it freezes out is also arbitrary. Therefore, true alignment of the disc pericentre with that of the binary is not captured in this 1D approach.

\section{Long-Term Mass Accretion and Net Angular Momentum Transfer}
\label{sec:longterm_mdot_jdot}

\begin{figure}
\begin{center}
\includegraphics[width=0.45\textwidth,clip]{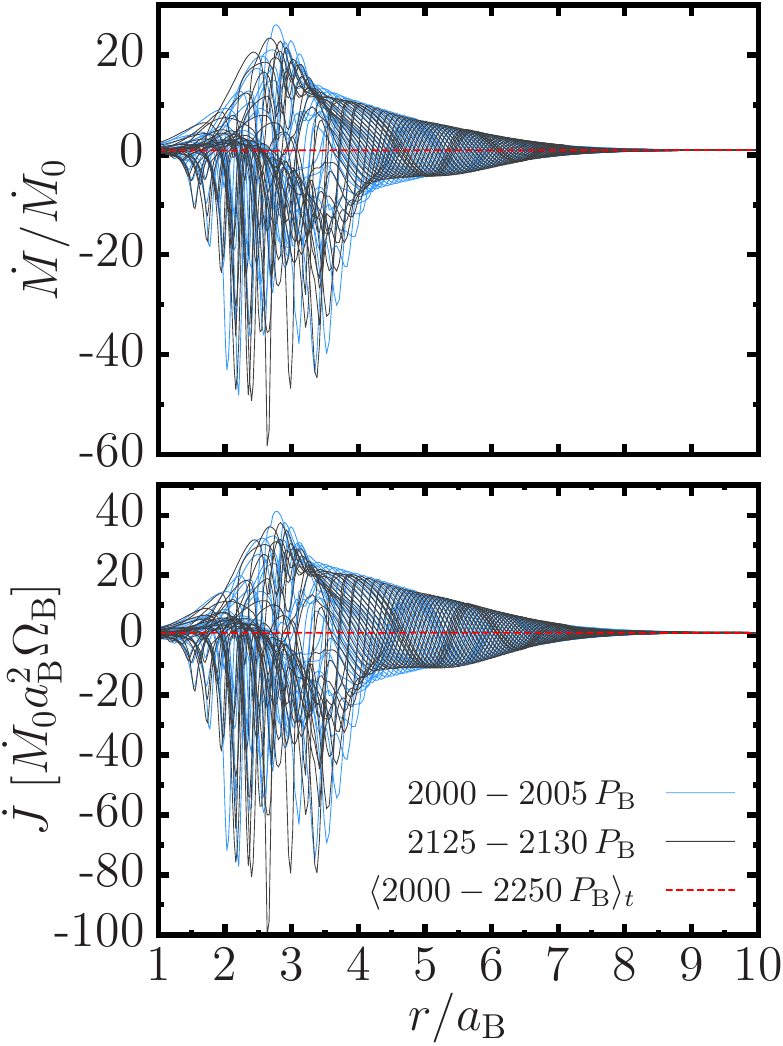}
\caption{Radial profiles of the mass accretion rate (top) and net angular momentum accretion rate (bottom) for the simulation with $q_\mathrm{B} = 1$, $e_\mathrm{B} = 0$, and $\alpha = 0.1$. The solid lines are instantaneous profiles, at $50$ different times, sampled over two $5 P_\mathrm{B}$ intervals $125 P_\mathrm{B}$ apart. The dashed lines are the time-averaged profiles, taken over $250 P_\mathrm{B}$. Although there are large fluctuations in these quantities both in time and radius, suitable time averaging results in remarkably flat profiles for both.}
\label{fig:mdot_jdot_variation}
\end{center}
\end{figure}

\begin{figure*}
\begin{center}
\includegraphics[width=0.99\textwidth,clip]{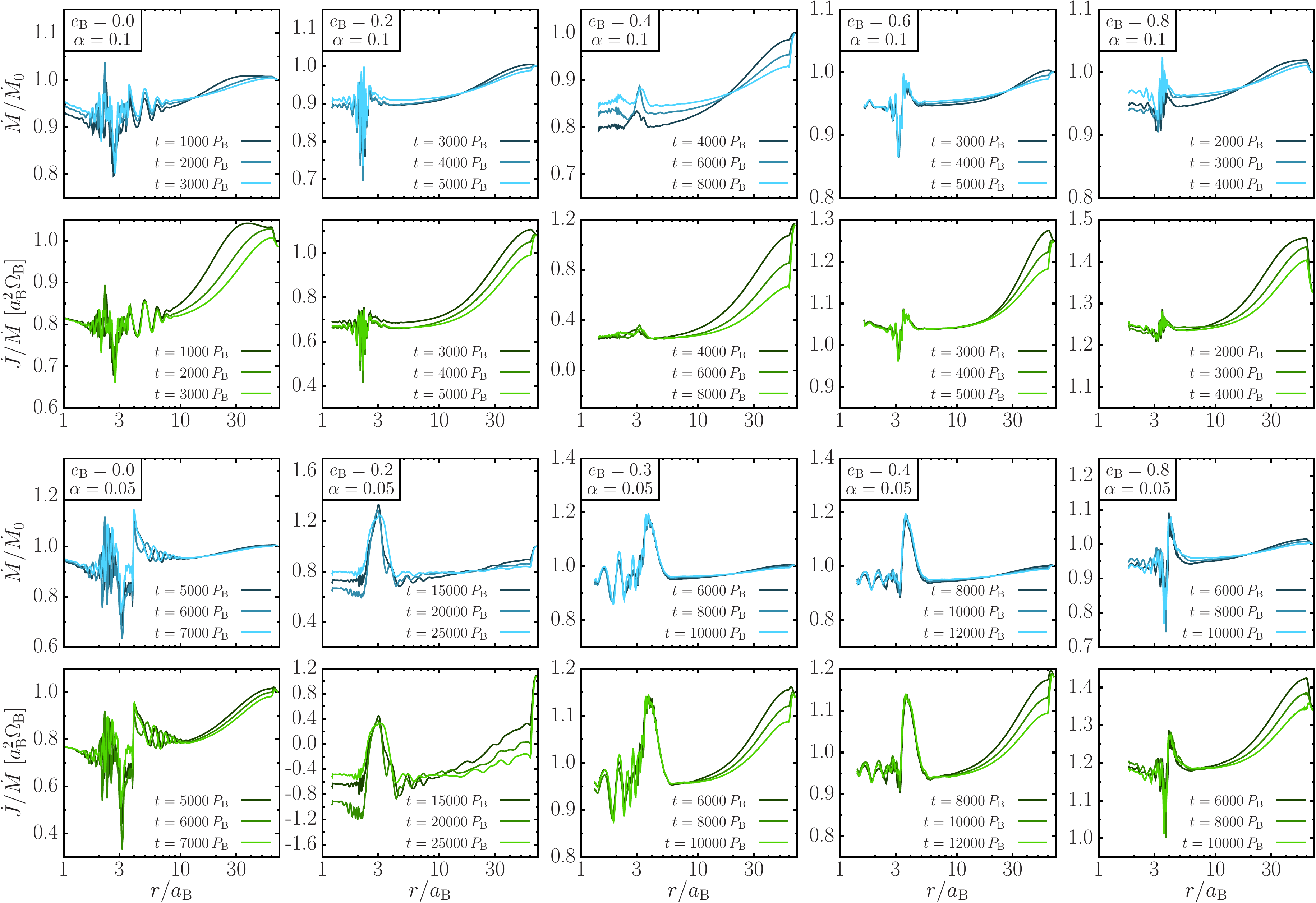}
\caption{Long-term time evolution of the profiles of the mass accretion rate, $\dot{M}$, and the ratio of the angular momentum and mass accretion rates, $\dot{J}/\dot{M}$, for equal mass binaries with different values of $e_\mathrm{B}$ and $\alpha$. Both $\dot{M}$ and $\dot{J}$ profiles are averaged over approximately one precession period [about $(300 - 600) P_\mathrm{B}$, for cases in which the disc precesses around the binary], or $50 P_\mathrm{B}$ (for cases in which the disc is apsidally locked to the binary), starting from the indicated time. The range of the $y$-axis varies between different panels, in order to focus on the residual radial variations in the profiles of the two quantities after time averaging. Note that the residual variations are much smaller than the instantaneous variations depicted in Fig.~\ref{fig:mdot_jdot_variation}. From the time-averaged profiles, we see that 1) $\dot{M}$ in the inner disc is very close to, but smaller than, the rate supplied at the outer boundary, and slowly approaches it with time, and 2) $\dot{J}/\dot{M}$ approaches a steady value in the inner region of the disc, and the size of this region grows with time. Together, these indicate that an ever-growing region of the inner disc is approaching a self-consistent quasi-steady state.}
\label{fig:mdot_jdot_evolution}
\end{center}
\end{figure*}

\begin{figure}
\begin{center}
\includegraphics[width=0.45\textwidth,clip]{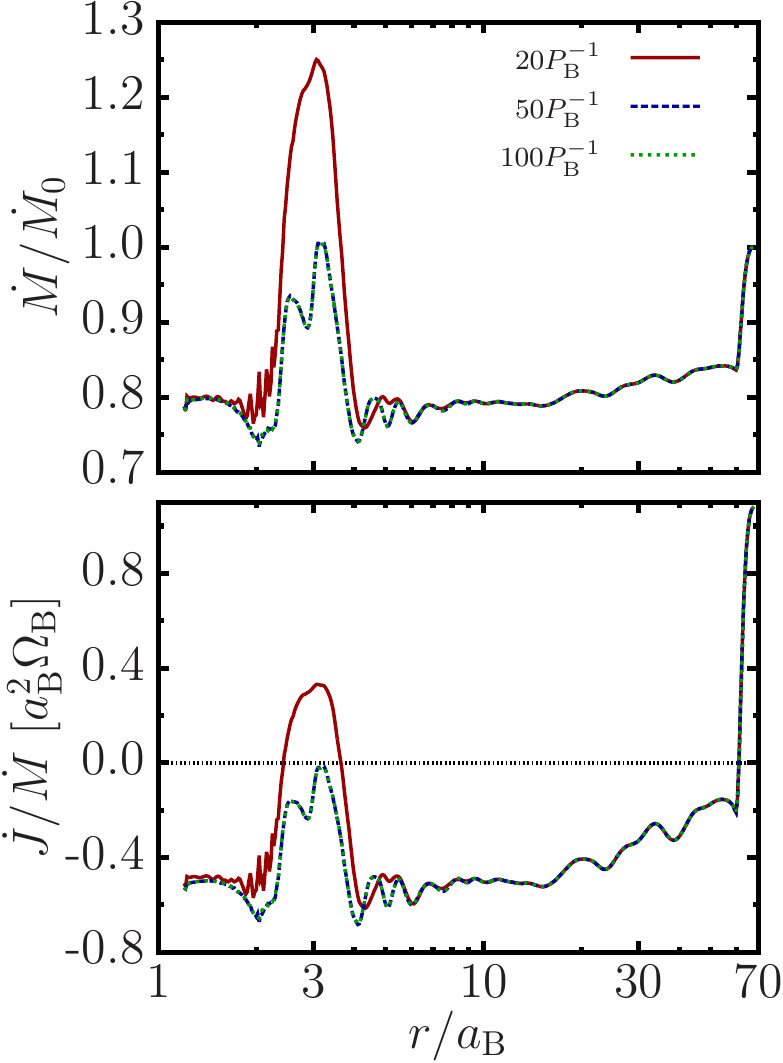}
\caption{Time averaged $\dot{M}$ and $\dot{J}$ for the simulation with $q_\mathrm{B} = 1.0, e_\mathrm{B} = 0.2, \alpha = 0.05$, each taken over $50$ binary orbits, with several time sampling rates, as indicated. Flatter profiles are obtained by using higher time sampling rates.}
\label{fig:mdot_jdot_time_sampling}
\end{center}
\end{figure}

\begin{figure}
\begin{center}
\includegraphics[width=0.45\textwidth,clip]{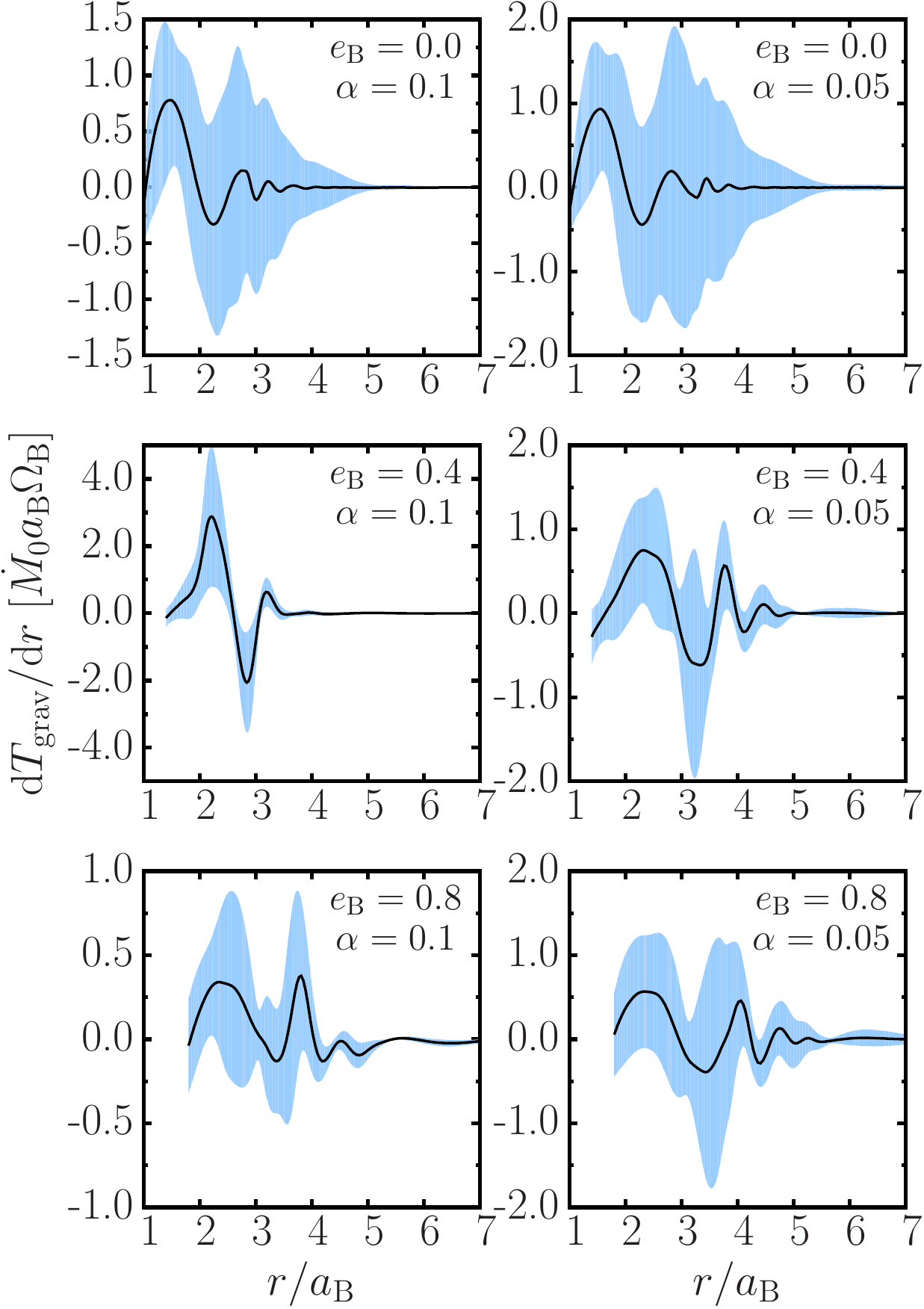}
\caption{Time averaged gravitational torque density $\mathrm{d}T_\mathrm{grav}/\mathrm{d}r$ (solid line), and its $1\sigma$ variations (shaded region), for equal mass binaries with several different eccentricities and two values of the disc $\alpha$. The torque density is an oscillatory function of $r$ and can be either positive or negative, and most of the torque is exerted on the disc within a radius of about $6 a_\mathrm{B}$.}
\label{fig:torque_density}
\end{center}
\end{figure}

\begin{figure}
\begin{center}
\includegraphics[width=0.45\textwidth,clip]{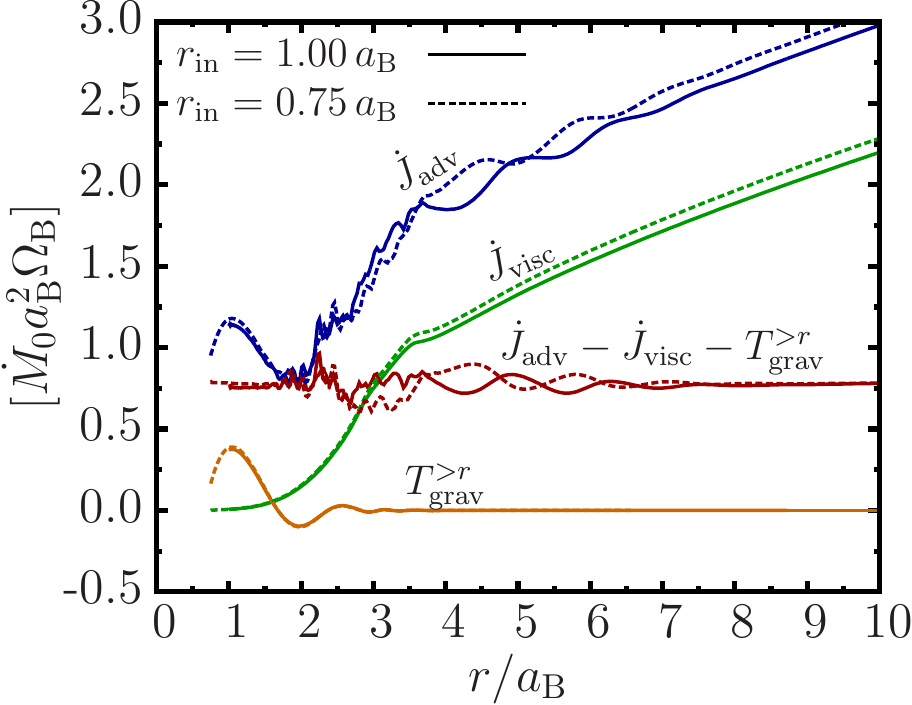}
\caption{Various contributions to the net angular momentum transfer rate as a function of radius for the simulation with $q_\mathrm{B} = 1$, $e_\mathrm{B} = 0$, and $\alpha = 0.1$. The net angular momentum transfer includes the advective, viscous, and gravitational contributions (see Eq.~\ref{eq:jdotdef}). The results for two different runs are shown: one with the standard inner boundary location, $r_\mathrm{in} = a_\mathrm{B}$ (solid lines), and one with an inner boundary located at $r_\mathrm{in} = 0.75a_\mathrm{B}$ (dashed lines). In both runs, all quantities are evaluated at around $2000 P_\mathrm{B}$, and averaged over $250 P_\mathrm{B}$.}
\label{fig:ang_mom_balance}
\end{center}
\end{figure}

\begin{figure}
\begin{center}
\includegraphics[width=0.45\textwidth,clip]{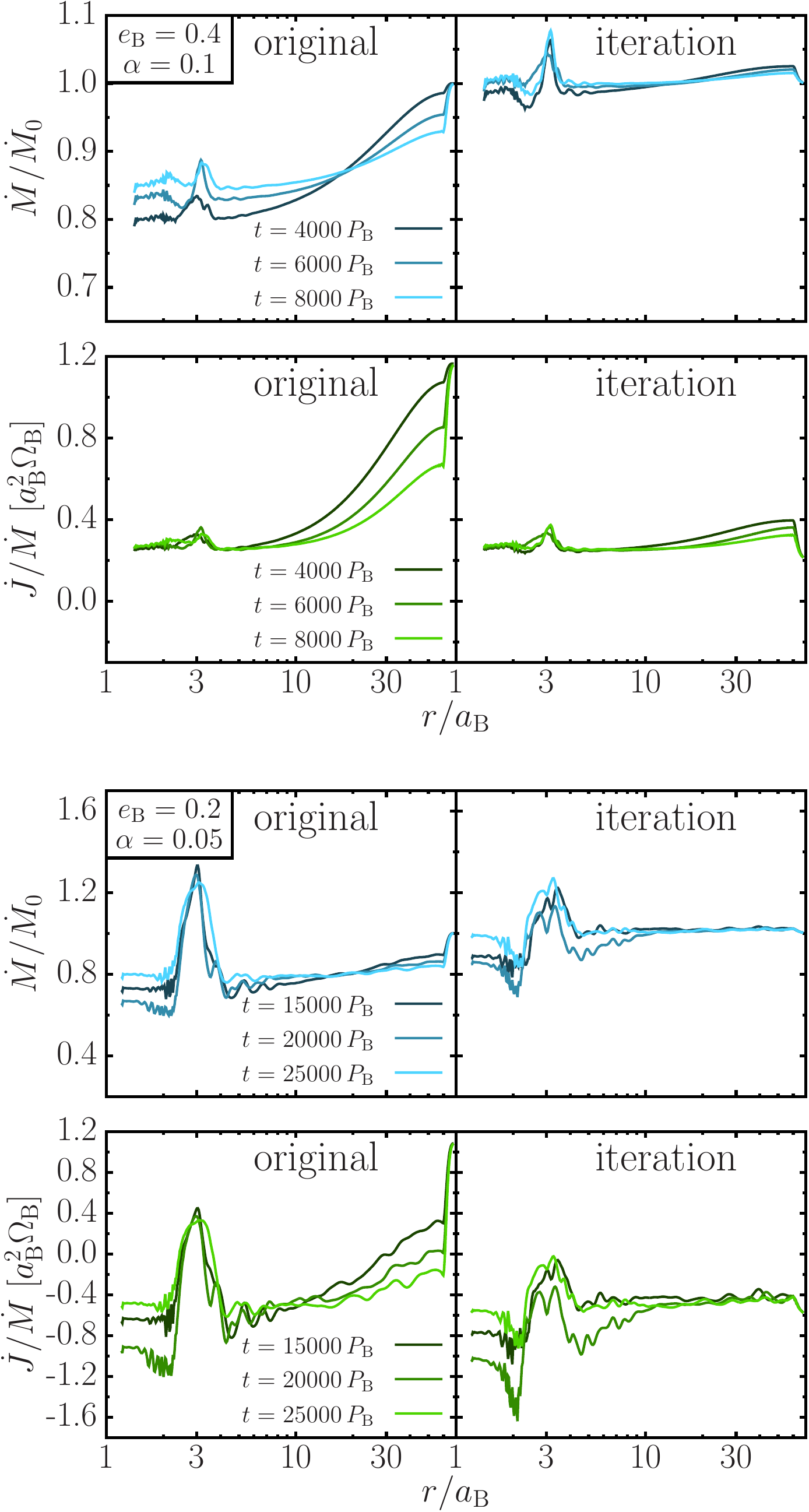}
\caption{Time averaged profiles of $\dot{M}$ and $\dot{J}/\dot{M}$ at several different times, as in Fig.~\ref{fig:mdot_jdot_evolution}, for two ``unusual'' cases (those which result in a minimum values of $\dot{J}/\dot{M}$ at the inner disc; see Fig.~\ref{fig:l_0}). These are shown for both the original runs (left), which use the standard initial conditions and outer boundary condition (see Eq.~\ref{eq:sigma_t0}), and for the iterated runs (right), for which the initial conditions and outer boundary condition are prescribed using the value of $l_0$ determined from the original run. Both are evolved for the same amount of time. In the iterated runs, both the $\dot{M}$ and $\dot{J}/\dot{M}$ profiles become flatter throughout the entire disc compared to the original runs, and the value of $\dot{M}$ in the inner disc is closer to the supply rate. This indicates that the value of $l_0 = \langle\dot{J}\rangle/\langle\dot{M}\rangle$ determined from the original run is close to the true global steady-state value for the disc.}
\label{fig:mdot_jdot_iteration}
\end{center}
\end{figure}

\begin{figure}
\begin{center}
\includegraphics[width=0.45\textwidth,clip]{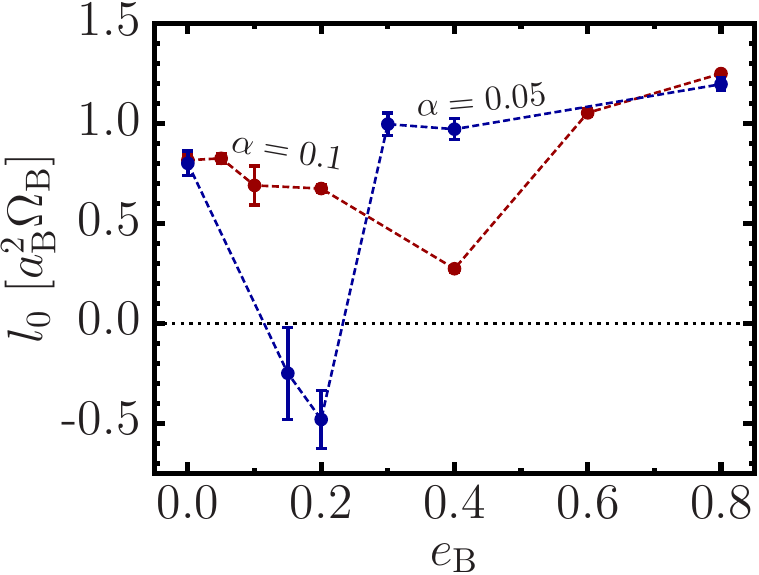}
\caption{Net angular momentum per unit mass received by the binary, $l_0 = \langle\dot{J}\rangle/\langle\dot{M}\rangle$, as a function of $e_\mathrm{B}$, for accretion onto equal-mass binaries, for two different values of the disc viscosity parameter $\alpha$. These values are determined by averaging the steady-state profiles of $\dot{J}/\dot{M}$ (see Fig.~\ref{fig:mdot_jdot_evolution}) from $r_\mathrm{in}$ to $10 a_\mathrm{B}$, and the error bars quantify the variations in these profiles. We see that $l_0$ is positive for most cases, except for when $\alpha = 0.05$ and $e_\mathrm{B} \sim 0.15 - 0.2$. The dip in the $l_0$ value at intermediate binary eccentricities ($0.4$ for the $\alpha =0.1$ case, and $0.2$ for the $\alpha=0.05$ case) corresponds to the case in which the inner eccentric disc is apsidally aligned with the binary (see Section \ref{sec:eccentricity_precession}).}
\label{fig:l_0}
\end{center}
\end{figure}

In this section, we use our simulations to determine the long-term net angular momentum transfer rate from the disc to the binary, taking into account mass accretion and gravitational torques. Given the large variabilities of the disc accretion on various time-scales (see Sections \ref{sec:short_timescale} and \ref{sec:eccentricity_precession}), it is essential that the simulations be carried out over sufficiently long times so that the disc reaches a true quasi-steady state in terms of mass accretion.

\subsection{Evolution of Global Mass Accretion Rate Profile}
\label{subsec:mdot_longterm}

Figure~\ref{fig:mdot_jdot_variation} (top panel) shows examples of both instantaneous radial profiles of $\dot{M}$ (computed using piecewise parabolic reconstruction of the fluid variables), as well as its time-averaged profile. In these examples, the time-averaged profile is computed over an interval of $250 P_\mathrm{B}$ (approximately the disc precession period). We see that the instantaneous $\dot{M}$ profiles exhibit large fluctuations (in both magnitude and sign), with an amplitude $20 - 60$ times larger than the mass supply rate $\dot{M}_0$. Nonetheless, the average $\dot{M}$ profile is very flat, such that its fluctuations are not visible when plotted on the same scale as the instantaneous profiles. Note that in Fig.~\ref{fig:mdot_jdot_variation}, some of the sharp peaks in the instantaneous profiles are captured by only a few grid points. In our high resolution runs, we find that the amplitude of these features are somewhat reduced, as is the amplitude of those in the averaged profiles. Our standard resolution runs typically achieve an accuracy of about $10$ per cent in the averaged $\dot{M}$.

We now focus on the small residual fluctuations in the time-averaged profiles of $\dot{M}$. Figure~\ref{fig:mdot_jdot_evolution} shows the results for $10$ different simulations, all with $q_\mathrm{B} = 1.0$, and different values of $e_\mathrm{B}$ and $\alpha$. The averaging is performed over either approximately one precession period, with a sampling rate of $5/P_\mathrm{B}$, or over $50 P_\mathrm{B}$, with a sampling rate of $20/P_\mathrm{B}$, for cases in which the inner discs are apsidally locked. We see that near the end of the simulations, many of the $\langle\dot{M}\rangle$ profiles have residual fluctuations with an amplitude of $30$ per cent or less, and in some cases, less than $10$ per cent. In the worst case, $(q_\mathrm{B},e_\mathrm{B},\alpha) = (1.0,0.2,0.05)$, the amplitude exceeds $50$ per cent. Some of this can be attributed to imperfect averaging, and the amplitude can be further reduced with a higher rate of time sampling. This is demonstrated in Fig.~\ref{fig:mdot_jdot_time_sampling}, in which the sampling rate is increased to $50/P_\mathrm{B}$, resulting in smaller fluctuations (about $30$ per cent in amplitude). No further advantage is gained by increasing the rate to $100/P_\mathrm{B}$. The $\langle\dot{M}\rangle$ profiles for other cases could also be made flatter with increased an time sampling rate. Thus, the amplitude of the fluctuations in Fig.~\ref{fig:mdot_jdot_evolution} can be considered an upper limit to the real residual fluctuations. We conclude that, at the end of each simulation, $\langle\dot{M}\rangle$ is constant with $r$ to within $30$ per cent (or much less), i.e., mass is flowing at an approximately steady rate throughout the disc.

We see from Fig.~\ref{fig:mdot_jdot_evolution} that, in general, at a given time, the time-averaged value of $\dot{M}$ at $r_\mathrm{in}$ is approximately equal to its value at $\sim 10 a_\mathrm{B}$. This value represents the average mass flow rate through the inner disc onto the binary. In the outer disc, $\dot{M}$ smoothly transitions to its value at $r_\mathrm{out}$, $\dot{M}_0$. At the end of each simulation, the inner disc $\langle\dot{M}\rangle$ is typically $80 - 90$ per cent of $\dot{M}_0$. The small difference between the inner disc $\langle\dot{M}\rangle$ and $\dot{M}_0$ is a consequence of the ``incorrect'' initial disc density profile: Eqs. \ref{eq:sigma_t0} and \ref{eq:ur_t0} assume that the specific angular momentum accreted onto the binary is $\sqrt{GM_\mathrm{B} r_\mathrm{in}}$. In Section \ref{sec:jdot_iteration}, we show that starting with better initial conditions (as determined by the angular momentum accretion rate) results in the accretion rate profile $\langle\dot{M}\rangle$ becoming indistinguishable from $\dot{M}_0$ in a much shorter time.

\subsection{Net Angular Momentum Transfer Rate to the Binary: Calculation and Result}

The rate of net angular momentum transfer across the disc, $\dot{J}$, is
\be
\label{eq:jdotdef}
\dot{J}(r,t) = \dot{J}_\mathrm{adv} - \dot{J}_\mathrm{visc} - T_\mathrm{grav}^{>r}
\ee
(see Appendix \ref{sec:ang_mom_cons}). The different terms contributing to $\dot{J}$ are the rate of change of angular momentum due to advection,
\be
\dot{J}_\mathrm{adv} = -\oint r^2\Sigma u_r u_\phi \mathrm{d}\phi,
\ee
the viscous torque,
\be
\dot{J}_\mathrm{visc} = -\oint r^3\nu\Sigma \left[\frac{\partial}{\partial r}\left(\frac{u_\phi}{r}\right) + \frac{1}{r^2}\frac{\partial u_\mathrm{r}}{\partial \phi} \right] \mathrm{d}\phi,
\ee
and the (integrated) gravitational torque,
\be
T_\mathrm{grav}^{>r} = \int_r^{r_\mathrm{out}}\frac{\mathrm{d}T_\mathrm{grav}}{\mathrm{d}r} \mathrm{d}r,
\ee
where
\be
\frac{\mathrm{d}T_\mathrm{grav}}{\mathrm{d}r} = -\oint r\Sigma \frac{\partial\Phi}{\partial\phi} \mathrm{d}\phi
\ee
is the gravitational torque density. In general, $\dot{J}$ is a function of $r$ and $t$. However, in a quasi-steady state state, its time-averaged value, $\langle\dot{J}\rangle$, is independent of $r$, so that angular momentum flows at a constant rate through the disc and onto the binary. The quasi-steady state is described by the eigenvalue
\be
\label{eq:l0def}
l_0 = \frac{\langle\dot{J}\rangle}{\langle\dot{M}\rangle},
\ee
which is the angular momentum per unit mass accreted onto the binary.  We now examine the evolution of $\langle\dot{J}\rangle$ towards a steady state, in order to compute the value of $l_0$ for each simulation. 

\subsubsection{Angular Momentum Transfer Rate Profile}

Figure~\ref{fig:mdot_jdot_variation} (lower panel) shows examples of the instantaneous and the time-averaged profiles of the net angular momentum transfer rate, $\dot{J}$. As in the case of the $\dot{M}$ profiles, the instantaneous fluctuations in $\dot{J}$ are very large (several orders of magnitude larger than the average value), but the fluctuations in the time-averaged profile are much smaller. In order to determine that the disc is evolving towards a quasi-steady state, we must examine these small residual fluctuations in $\dot{J}$.

We first look at the contribution to the angular momentum transfer due to the gravitational torque from the tidal potential of the binary. Although the radially-integrated gravitational torque is the quantity which contributes to $\dot{J}$, it is illustrative to inspect its radial derivative, the gravitational torque density $\mathrm{d}T_\mathrm{grav}/\mathrm{d}r$. Examples of its time-averaged profiles and temporal variations are shown in Fig.~\ref{fig:torque_density}. We see that the magnitude of the torque density is only appreciable for $r \lesssim 6 a_\mathrm{B}$. The torque density is an oscillatory function of $r$, and can change sign. Thus, it resembles the wave function of an isolated Lindblad resonance (e.g., Meyer-Vernet \& Sicardy 1987), or the combination of a few resonances. This type of torque density profile is different from the one arising from the superposition of many Lindblad resonances [e.g., in the impulse approximation, $\mathrm{d}T_\mathrm{grav}/\mathrm{d}r \propto 1/(r-a_\mathrm{B})^4$; see Lin \& Papaloizou 1979], which is only relevant for small mass ratios, and not the mass ratios considered in our simulations ($q_\mathrm{B}= 0.2 - 1.0$).

We now look at the different contributions to $\dot{J}$. Figure~\ref{fig:ang_mom_balance} shows examples of the time-averaged profiles of the three terms on the right-hand side of Eq.~(\ref{eq:jdotdef}), which are due to advection, viscous torque, and gravitational torque (the latter being the integral of the quantity shown in Fig.~\ref{fig:torque_density}), as well as the net angular momentum transfer rate, $\dot{J}$, for two different simulations. Both simulations have the same binary and disc parameters, but one uses the standard inner boundary location adopted in the rest of our simulations ($r_\mathrm{in}= a_\mathrm{B}$ in this case), while the other uses a slightly smaller inner boundary radius, $r_\mathrm{in} = 0.75 a_\mathrm{B}$ (in this simulation, a larger number of radial grid cells were used so that both runs have the same spatial resolution).  In both simulations, the profiles of each of the contributions to $\dot{J}$ are very similar, and we find the same nearly constant value of the total $\langle\dot{J}\rangle$ in the inner disc. Thus, an equilibrium value $\langle\dot{J}\rangle$ is reached the inner disc, and it does not depend on the location of the inner boundary.

Figure~\ref{fig:mdot_jdot_evolution} shows the $\langle\dot{J}\rangle/\langle\dot{M}\rangle$ profiles at different times for $10$ different simulations. The fluctuations in these averaged profiles are much smaller than in the instantaneous profiles depicted in Fig.~\ref{fig:mdot_jdot_variation}. At the end of each simulation, $\langle\dot{J}\rangle/\langle\dot{M}\rangle$ is nearly constant in the inner part of the disc ($r \lesssim 10 a_\mathrm{B}$). The fluctuations around the average are typically $25$ per cent, but in some cases are less than $10$ per cent. In the ``worst'' case, $(q_\mathrm{B}, e_\mathrm{B}, \alpha) = (1.0,0.2,0.05)$, the fluctuations are of order unity, although some of this can be attrubuted to the time sampling rate (see Fig.~\ref{fig:mdot_jdot_time_sampling}). For this reason, we ran this simulation for much longer than others ($25000 P_\mathrm{B}$), in order to ensure relaxation of the inner disc. In all cases, $\langle\dot{J}\rangle/\langle\dot{M}\rangle$ transitions at large radii ($r > 10 a_\mathrm{B}$) to its prescribed initial value at $r_\mathrm{out} = 70 a_\mathrm{B}$, corresponding to $\dot{J}(r_\mathrm{out}) = \dot{M_0} \sqrt{GM_\mathrm{B}r_\mathrm{in}}$ (see Eq.~\ref{eq:sigma_t0}).

\subsubsection{Iterated Runs}
\label{sec:jdot_iteration}

Since $\dot{J}(r_\mathrm{out})$ is fixed, and since the damping zone near the outer boundary relaxes $\dot{J}$ to this value, $\langle\dot{J}\rangle$ in the outermost part of the disc is never equal to $\langle\dot{J}\rangle(r_\mathrm{in})$, which is computed self-consistently. If we were to run the simulation for a viscous time-scale at the boundary of the damping zone, $r_\mathrm{damp} = 60 a_\mathrm{B}$ ($\sim 3 \times 10^4 P_\mathrm{B}$ for $\alpha = 0.1$), we would see $\langle\dot{J}\rangle$ take on a constant value for $r_\mathrm{in} < r < r_\mathrm{damp}$, and transition from this value to $\dot{J}(r_\mathrm{out})$ in the damping zone. To produce a true steady state, in which $\langle\dot{J}\rangle$ is constant for all $r$, an iterative process is required. In this process, we use the value of $\langle\dot{J}\rangle$ to construct the initial conditions and outer boundary condition of a new simulation [by replacing $(r_\mathrm{in}/r)^{1/2}$ in Eq.~(\ref{eq:sigma_t0}) with $l_0/\sqrt{GM_\mathrm{B} r}$], so that a steady state characterized by $l_0$ is imposed in the outer disc. Once the inner disc has viscously relaxed, we can achieve a global quasi-steady state, which has $\langle\dot{J}\rangle(r_\mathrm{in}) \approx \dot{J}(r_\mathrm{out})$. We performed this iteration for two cases: $(q_\mathrm{B},e_\mathrm{B},\alpha) = (1.0,0.4,0.1)$ and $(1.0,0.2,0.05)$. These are cases are of particular interest, as they have the lowest values of $l_0$ (for their corresponding values of $q_\mathrm{B}$ and $\alpha$).

The results of the iterated simulations are shown in Fig.~\ref{fig:mdot_jdot_iteration}. In the original runs, $\langle\dot{M}\rangle$ at $r_\mathrm{in}$ evolves slowly towards $\dot{M}_0$, reaching a value of $\sim 0.8 \dot{M}_0$ at the end of the simulation, for both simulations shown. In the iterated runs, $\langle\dot{M}\rangle$ at $r_\mathrm{in}$ reaches a value very close to $\dot{M}_0$ much sooner, since the initial conditions more closely match the true quasi-steady state of the disc. In the inner disc, $\langle\dot{J}\rangle/\langle\dot{M}\rangle$ has a value close to the one in the original runs, but it is much closer to being constant throughout the entire disc, and the transition to the outer boundary value is less extreme. Thus, the self-consistent angular momentum transfer rate calculated for the the inner disc is independent of the artificially imposed value at the outer boundary, and represents the true quasi-steady state of the disc. It is therefore not necessary to perform this iteration for all of our simulations, as it does not affect the profile of $\langle\dot{J}\rangle/\langle\dot{M}\rangle$ in the inner disc (i.e., the value of $l_0$).

\subsubsection{Value of the Eigenvalue $l_0$}

We have shown that nearly constant profiles of $\langle\dot{J}\rangle$ and $\langle\dot{M}\rangle$, representing a quasi-steady state, are achieved in the inner disc at the end of our simulations. Thus, we now compute the value of the eigenvalue which characterizes the quasi-steady state, $l_0$, for each simulation.  In a perfect quasi-steady state, in which $\langle\dot{J}\rangle$ and $\langle\dot{M}\rangle$ are independent of $r$, $l_0$ is given by Eq.~(\ref{eq:l0def}). Since, in our simulations, $\langle\dot{J}\rangle$ and $\langle\dot{M}\rangle$ are not truly independent of $r$, we compute the value of $l_0$ by taking the radial average of $\langle\dot{J}\rangle/\langle\dot{M}\rangle$ between $r_\mathrm{in}$ and $r_\mathrm{cut} = 10 a_\mathrm{B}$,
\be
l_0 = \frac{1}{r_\mathrm{cut}-r_\mathrm{in}} \int_{r_\mathrm{in}}^{r_\mathrm{cut}} \frac{\langle\dot{J}\rangle(r)}{\langle\dot{M}\rangle(r)} \mathrm{d}r,
\ee
using $\langle\dot{J}\rangle$ and $\langle\dot{M}\rangle$ at the end of each simulation. We also compute the standard deviation of $\langle\dot{J}\rangle/\langle\dot{M}\rangle$ over the same radial interval,
\be
\Delta l_0 = \left\{\frac{1}{r_\mathrm{cut}-r_\mathrm{in}} \int_{r_\mathrm{in}}^{r_\mathrm{cut}} \left[\frac{\langle\dot{J}\rangle(r)}{\langle\dot{M}\rangle(r)} - l_0\right]^2 \mathrm{d}r\right\}^{1/2},
\ee
in order to quantify the systematic uncertainty in $l_0$.

The results are given in Table \ref{tab:summary} and shown in Fig.~\ref{fig:l_0} for simulations with equal mass binaries (the values of $l_0$ for the rest of our simulations are only given in Table \ref{tab:summary}). We see that $l_0$ almost always has a positive value\footnote{Figure~\ref{fig:l_0} contains an additional point, corresponding to the case $(e_\mathrm{B}, \alpha) = (0.15, 0.05$), which is not analysed in detail in the other numerical results sections. This case, which was evolved for $10000 P_\mathrm{B}$, was found to have a negative value of $l_0$. It serves to demonstrate that the calculation of $l_0$ for the case $(e_\mathrm{B}, \alpha) = (0.2, 0.05$), which would otherwise be the only occurrence of a negative $l_0$, is not spurious.}. Thus, on average, the binary gains net angular momentum. For $e_\mathrm{B} = 0$, the value of $l_0$ (about $0.80$, in units of $a_\mathrm{B}^2\Omega_\mathrm{B}$) is the same to within our uncertainty for both $\alpha = 0.1$ and $0.05$. As $e_\mathrm{B}$ increases, $l_0$ first decreases, reaching a minimum ($0.27$ for $\alpha = 0.1$ and $-0.48$ for $\alpha = 0.05$), before increasing again. At $e_\mathrm{B} = 0.8$, the values of $l_0$ are almost the same for both values of $\alpha$ ($1.25$ for $\alpha = 0.1$ and $1.20$ for $\alpha = 0.05$). For both values of $\alpha$, the minimum value of $l_0$ occurs at the value of $e_\mathrm{B}$ for which the disc is apsidally locked with the binary (see Section \ref{sec:eccentricity_precession}).

\subsection{Net Angular Momentum Transfer Rate to the Binary: Discussion}
\label{subsec:l_0_discussion}

Figure \ref{fig:l_0} represents the most important result of this paper. We find that the (long-term averaged) specific angular momentum received by the binary, $l_0=\langle\dot J\rangle/\langle\dot M\rangle$, is positive for most binary eccentricities (including $e_\mathrm{B}=0$). This directly contradicts the previous numerical results (e.g., MacFadyen \& Milosavljevi\'{c} 2008), as well as the the assumption adopted in many papers on the disc-driven merger of supermassive black hole binaries (e.g., Armitage \& Natarajan 2002; Haiman et al.~2009; Chang et al.~2010).

As we have emphasized above, $l_0$ is an eigenvalue of the accretion flow and can only be determined by the global solution of the flow with a proper treatment of the boundaries. In our simulations we have only considered two values of the viscosity parameter ($\alpha=0.1$ and $0.05$) and the disc aspect ratio $H/r=0.1$. Can a smaller viscosity qualitatively change our result? (For example, a binary surrounded by a ``non-accreting'' disc would lose angular momentum to the disc through gravitational torque.) We think this is unlikely. Indeed, the angular momentum flux across the disc can be written as (see Eq.~\ref{eq:jdotdef})
\be
\dot J(r,t)=\dot M l -3\pi\nu\Sigma l - \Sigma l F(r),
\ee
where we have used the fact that the gravitational torque is proportional to the surface density, and thus $T_{\rm grav}^{>r}=\Sigma l F(r)$, with $F(r)>0$.  In steady state, $\dot J=\dot M l_0$, and the disc surface density is given by 
\be
\label{eq:sigma_balance}
\Sigma ={{\dot M}(1-l_0/l)\over 3\pi\nu +F(r)}.
\ee
Thus, while a reduced viscosity indeed increases the surface density, it does not necessarily change the balance between the advective, viscous, and gravitational angular momentum fluxes.

Among previous numerical studies, the work of MacFadyen \& Milosavljevi\'{c} (2008) was the most similar to ours. These authors considered $H/r=0.1$ but a lower disc viscosity ($\alpha=0.01$).  (They also adopted a polar grid in the domain between $r_{\rm in}=a_\mathrm{B}$ and $r_{\rm out}=100a_\mathrm{B}$.)  With such a small viscosity, the ``viscous relaxation'' radius (see Eq.~\ref{eq:r_relax}) at $t=4000P_\mathrm{B}$ (the typical duration of their runs) is only about $3a_\mathrm{B}$. Moreover, their initial surface density profile is far from the steady state even for $r \gg r_{\rm in}$. Thus, we suggest that their findings concerning the reduction of mass accretion onto the binary and the dominance of the gravitational torque relative to advective torque (therefore a negative $l_0$) reflected only the ``transient'' phase of their simulations. Several more recent numerical studies (e.g., D'Orazio et al.~2013; Farris et al.~2014) have explored various aspects of circumbinary accretion, but did not examine the detailed balance of angular momentum transfer.

Shi et al.~(2012) obtained a positive value of $l_0$ in their MHD simulations of circumbinary discs. However, as the duration of their simulations is only $\sim 100 P_\mathrm{B}$ (due to the costly 3D numerical method), it is unlikely that a quasi-steady state was reached, so their value of $l_0$ may not properly characterize the long-term evolution of the binary and disc. The magnitude of their calculated $l_0$ is not large enough to cause orbital expansion, in contrast to our results (see Section \ref{subsubsec:orbital_evolution}).

A most striking feature of Fig.~\ref{fig:l_0} is that $l_0$ is, in general, a non-monotonic function of $e_\mathrm{B}$. For each value of $\alpha$, the minimum of $l_0$ occurs at the binary eccentricity ($e_\mathrm{B}=0.4$ for $\alpha=0.1$ and $e_\mathrm{B}=0.2$ for $\alpha=0.05$) for which the inner disc becomes apsidally aligned with the binary (see Section \ref{sec:eccentricity_precession}, particularly Figs.~\ref{fig:power_lowfreq}-\ref{fig:ecc_vector}). At both lower and higher binary eccentricities, the inner disc precesses coherently, giving rise to a larger value of $l_0$. We can only speculate that this remarkable inter-dependence of the long-term disc variability and the global disc eigenvalue $l_0$ is the result of the intricate balance between the different contributions to $\dot{J}$ and the details of the accretion dynamics.

In this paper we have only considered binaries with equal masses or modest mass ratios ($q_\mathrm{B}\ge 0.2$), for which we have found $l_0$ is mostly positive (see Table \ref{tab:summary}). Our simuations do not apply to binaries with more extreme mass ratios, for which an annular gap is opened by the (small-mass) secondary instead of a common cavity surrounding both components of the binary (D'Orazio et al.~2016). For such extreme mass ratios, the binary may lose angular momentum to the disc, analogous to Type II planet migration (e.g., Lin \& Papaloizou 1986). Another caveat of our study is that we have neglected the self-gravity of the disc, which may play a role in the angular momentum transfer to the binary (e.g., Cuadra et al.~2009). This amounts to assuming that the disc mass $\pi r^2 \Sigma$ (inside a few $a_\mathrm{B}$'s) is much less than the mass of the binary $M_\mathrm{B}$. Binary black holes embedded in a massive nuclear disc may suffer angular momentum loss to the disc due to dynamical friction (e.g., Escala et al.~2005).

\section{Summary and Discussion}
\label{sec:discussion}

\subsection{Summary of Results}
We have carried out long-term, two-dimensional simulations of viscous circumbinary discs, for a range of binary eccentricities, mass ratios, and viscosity parameters. We have focused on the scenario in which mass is supplied to the outer disc at a fixed rate, and examined the quasi-steady state of the disc, as characterized by a steady flow of mass and angular momentum throughout the disc. Our key results are:

\begin{itemize}

\item The structure of the inner disc is characterized by the radius $r_\mathrm{peak} = (3 - 5) a_\mathrm{B}$ (where $a_\mathrm{B}$ is the binary semi-major axis), at which the azimuthally-averaged surface density reaches a maximum, and another radius, $r_{0.1} = (1.6 - 2.6) a_\mathrm{B}$, at which the surface density falls to $10$ per cent of its value at $r_\mathrm{peak}$. While $r_\mathrm{peak}$ is non-monotonic in $e_\mathrm{B}$, $r_{0.1}$ strictly increases with $e_\mathrm{B}$ (see Fig. \ref{fig:truncation}), and is a good representation of the ``truncation radius'' of the disc, as it agrees with the theory of inner disc truncation due to the clearing of a gap at increasingly higher order Lindblad resonances (as a function of $e_\mathrm{B}$) to within $20$ per cent.

\item The mass accretion onto the central binary is highly variable (see Fig.~\ref{fig:mdot_power}), with the dominant period of variability either equal to the binary period $P_\mathrm{B}$ or about $5 P_\mathrm{B}$ (``low-frequency variability''). The low-frequency variability occurs for $e_\mathrm{B} \lesssim 0.05$ and $q_\mathrm{B} \gtrsim 0.2$, and is associated with the Keplerian motion of density ``lumps'' that are continuously formed and destroyed near the inner edge of the disc (slightly interior to $r_\mathrm{peak}$). The accretion variability at $P_\mathrm{B}$ occurs for $e_\mathrm{B}\gtrsim 0.05$, in which case no lump is formed at the inner disc (see Fig.~\ref{fig:lump}).

\item The inner region of the disc ($3 a_\mathrm{B} \lesssim r \lesssim 6 a_\mathrm{B}$) is generally eccentric. For both low ($e_\mathrm{B} \lesssim 0.2$) and high ($e_\mathrm{B} \gtrsim 0.4$) binary eccentricities, the eccentric disc coherently precesses around the binary with a period of several hundred binary orbits. For intermediate eccentrities ($e_\mathrm{B} = 0.2 - 0.4$, depending on the disc viscosity parameter), the eccentric disc instead becomes apsidally locked with the binary (see Figs.~\ref{fig:power_lowfreq}-\ref{fig:ecc_vector}). The behavior of the inner eccentric disc may be explained in a linear, one-dimensional framework, using a combination of eccentricity excitation at multiple eccentric Lindblad resonances, the strengths of which depend on $e_\mathrm{B}$, and viscous eccentricity damping (see Section \ref{sec:eccentricity_precession_theory}).

\item Although the ``instantaneous'' mass accretion rate across the disc is highly variable (see Fig.~\ref{fig:mdot_jdot_variation}), the time-averaged accretion rate (averaged over hundreds of $P_\mathrm{B}$, corresponding to the precession period of the inner eccentric disc), is constant across the disc, and approximately equals  the mass supply rate at the outer radius of the disc (see Fig.~\ref{fig:mdot_jdot_evolution}). Although this result contradicts several previous claims, it does agree with the recent simulations using the moving mesh code \textsc{arepo} (Mu\~{n}oz \& Lai 2016) that resolve accretion onto individual components of the binary. 

\item Our most important finding concerns the time-averaged  angular momentum transfer rate from the disc to the binary. This angular momentum transfer  includes contributions from mass accretion, viscous stress and gravitational torque between the binary and the disc. We find that, in the quasi-steady state, the specific angular momentum transferred to the binary, $l_0=\langle\dot{J}\rangle/\langle\dot{M}\rangle$, depends on the binary eccentricity in a non-monotonic manner (Fig.~\ref{fig:l_0}; see Table \ref{tab:summary} for a list of the computed $l_0$ values). Contrary to many previous claims and assumptions, we find that $l_0$ is positive for most $e_\mathrm{B}$'s, implying that the binary receives net angular momentum. The minimum $l_0$ occurs at intermediate binary eccentricities (around $0.2-0.4$, depending on the viscosity parameter), and corresponds to the regime where the inner eccentric disc is apsidally aligned with the binary (see Section \ref{subsec:l_0_discussion} for discussion).

\end{itemize}

\subsection{Astrophysical Implications}

\begin{figure}
\begin{center}
\includegraphics[width=0.45\textwidth,clip]{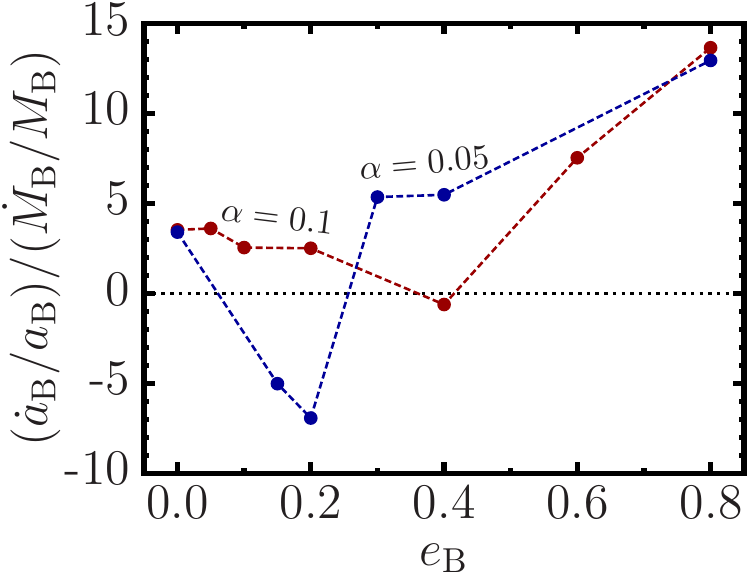}
\caption{The rate at which the binary semi-major axis changes, $\dot{a}_\mathrm{B}/a_\mathrm{B}$ (Eq.~\ref{eq:adot}), normalized by the rate at which its mass grows, $\dot{M}_\mathrm{B}/M_\mathrm{B}$, as inferred from the values of $l_0$ shown in Fig.~\ref{fig:l_0}. We have assumed that the members of the equal mass binary take on equal shares of the accreted mass, and that the binary eccentricity does not change. Under these assumptions, the semi-major axis of the binary grows with time for a wide range of $e_\mathrm{B}$.}
\label{fig:adot}
\end{center}
\end{figure}

\subsubsection{Binary Orbital Evolution}
\label{subsubsec:orbital_evolution}

As noted before (Section \ref{sec:introduction} and Section \ref{subsec:l_0_discussion}), binary-disc interaction has long been suggested to play an important role in driving the orbital decay of supermassive black hole (SMBH) binaries. The rate of change of the binary angular momentum is related to the rate of change of its orbital elements, $\dot{a}_\mathrm{B}$ and $\dot{e}_\mathrm{B}$, as well as the mass accretion rate of the individual stars, $\dot{M}_1$ and $\dot{M}_2$, according to
\be
\frac{\dot{J}_\mathrm{B}}{J_\mathrm{B}} = \frac{\dot{M}_1}{M_1} + \frac{\dot{M}_2}{M_2} - \frac{1}{2}\frac{\dot{M}_\mathrm{B}}{M_\mathrm{B}} + \frac{1}{2}\frac{\dot{a}_\mathrm{B}}{a_\mathrm{B}} - \left(\frac{e_\mathrm{B}}{1-e_\mathrm{B}^2}\right)\dot{e}_\mathrm{B}.
\ee
We cannot track the mass accretion rates of the individual stars in the simulations presented in this paper, but for an equal mass binary, we may assume that on average, $\dot{M}_1 = \dot{M}_2 = \dot{M}_\mathrm{B}/2$, due to symmetry (this is true on average, but the symmetry can be temporarily broken in an alternating fashion due to precession, e.g., Dunhill et al. 2015; Mu\~{n}oz \& Lai 2016). Solving the full evolution of the binary orbit also requires knowledge of the rate at which it gains or loses energy, the determination of which is beyond the scope of this work. If instead we simply assume that $\dot{e}_\mathrm{B} = 0$, the orbital evolution is described by
\be
\label{eq:adot}
\frac{\dot{a}_\mathrm{B}}{a_\mathrm{B}} = 8\left(\frac{l_0}{l_\mathrm{B}}-\frac{3}{8}\right)\frac{\dot{M}_\mathrm{B}}{M_\mathrm{B}},
\ee
where $l_\mathrm{B} = [GM_\mathrm{tot}a_\mathrm{B}(1-e_\mathrm{B}^2)]^{1/2}$ is the specific angular momentum of the binary. Under this assumption, the sign of $\dot{a}_\mathrm{B}/a_\mathrm{B}$ is determined by whether $l_0$ is larger or smaller than $3l_\mathrm{B}/8$.  Figure \ref{fig:adot} shows the orbital evolution rate for equal-mass binaries, using the values of $l_0$ determined in our numerical simulations (see Table \ref{tab:summary} and Fig.~\ref{fig:l_0}). Typically, $\dot{a}_\mathrm{B}$ is positive, meaning that the binary semi-major axis grows with time. We find two cases for which the binary shrinks, both corresponding to the minimum of $l_0(e_\mathrm{B})$, which is also associated with apsidal alignment of the eccentric disc and binary. We reiterate that Eq.~\ref{eq:adot} applies only when $\dot{e}_\mathrm{B} = 0$. Eccentricity damping ($\dot{e}_\mathrm{B}<0$) would reduce $\dot{a}_\mathrm{B}$.

This result is contrary to the common assumption that interactions with a surrounding disc lead to binary shrinkage (e.g., Begelman et al. 1980; Armitage \& Natarajan 2002; MacFadyen \& Milosavljevi\'{c} 2008). Therefore, at least for binary black holes with comparable masses, circumbinary discs may not provide a solution to the ``final parsec problem'' for SMBH binaries. As a caveat, we reiterate that the simulations in this paper were restricted to binaries with equal masses or modest mass ratios, and we have also assumed that the inner disc (within $\sim 10 a_\mathrm{B}$) has a negligible mass compared to the binary (see Section \ref{subsec:l_0_discussion}). More systematic studies of binaries with more extreme mass ratios and massive discs are warranted.

\subsubsection{Circumbinary Planet Formation}

Circumbinary protoplanetary discs, a number of which have been observed (e.g., Dutrey et al.~1994; Mathieu et al.~1997; Simon et al.~2000), are the birthplaces of circumbinary planets, such as those discovered by the \textit{Kepler} mission (e.g., Doyle et al. 2011; Welsh et al. 2012; Orosz et al. 2012). Many of these planets have semi-major axes a few times larger than that of the binary, and are therefore close to the dynamical stability limit (Holman \& Wiegert 1999). This radius is also close to the inner truncation radius of the disc, which may play a role in halting planetary migration (e.g., Nelson 2003; Pierens \& Nelson 2007; Kley \& Haghighipour 2014, 2015). 

The disc truncation radius increases with binary eccentricity (see Fig.~\ref{fig:truncation}). The gradients of the surface density profile in the inner disc are also sensitive to the binary eccentricity, becoming steepest when it takes on intermediate values. These steep density profiles may affect planet migration in these regions.

The growth of planetesimals may be inefficient close to the binary as a result of its gravitational perturbations (e.g., Moriwaki \& Nakagawa 2004; Paardekooper et al. 2012; Meschiari 2012; Rafikov 2013; Silsbee \& Rafikov 2015). The eccentricity of planetesimals embedded in the disc is damped due to gas drag, but it cannot reach zero, since the finite eccentricity of the disc itself serves as a lower limit. The average relative velocity between planetesimals is approximately $v_\mathrm{rel} \approx e v_\mathrm{K}$, where $v_\mathrm{K}$ is the Keplerian orbital velocity. At a separation of $(3-4) a_\mathrm{B}$ from the binary, near the dynamical stability limit where many of the Kepler circumbinary planets reside, we find that the disc has an eccentricity of $0.05 - 0.2$, which for a binary separation of $0.2 \mathrm{AU}$, corresponds to relative planetesimal velocities of several $\mathrm{km}/\mathrm{s}$. As this is much larger than their escape velocities (e.g., a few $\mathrm{m}/\mathrm{s}$ for $10 \mathrm{km}$ bodies), collisional growth of planetesimals is likely to be strongly inhibited in this region. However, we find that for intermediate binary eccentricities, the eccentric disc becomes apsidally aligned with the binary. If, in these cases, planetesimal orbits closely follow those of fluid elements in the disc (due to the effects of gas drag), then the planetesimals may be apsidally aligned with one another. As a result, they may have much smaller relative velocities, allowing collisional growth to occur. A full treatment of both the gas and planetesimal dynamics is required to assess the plausibility of this outcome.

\subsection{Limitations and Prospects}

Throughout this work, we have studied the accretion flow around binaries while discarding the region of the computational domain where individual circumsingle discs would form around each component of the binary. This simplification makes the problem tractable in cylindrical-coordinate grids. In addition, ignoring the inner region makes the long-term integrations presented here feasible, as it directly imposes a lower limit on the shortest computational time-step. However, as a consequence, the details of accretion onto the individual members of the binary -- such as mass and momentum transfer --  are lost. Simulating this co-orbital region is key to understanding the mass-growth and orbital evolution of the central binary, but it represents a significant computational challenge. Long-term simulation of individual accretion in binaries while retaining the features of \textsc{fargo}-like finite-volume codes such as \textsc{pluto} has only been accomplished recently (Farris 2014, Mu{\~n}oz \& Lai 2016). These simulations are now possible thanks to the implementation of novel meshing algorithms in the ground-breaking codes \textsc{disco} (Duffell \& MacFadyen 2013, Duffell 2016) and \textsc{arepo} (Springel 2010, Mu{\~n}oz et al. 2014). In future work (Mu\~{n}oz, Miranda \& Lai, in prep), we will complement the results of our \textsc{pluto} simulations with \textsc{arepo} simulations. As \textsc{arepo} and \textsc{pluto} are both finite-volume schemes (albeit with different-order reconstruction), the cell-centred primitive variables of a \textsc{pluto} snapshot can be directly mapped onto an \textsc{arepo} unstructured grid simulation and then evolved onwards. With virtually identical initial conditions, the mass and angular momentum exchange between the disc and the binary can be contrasted for these two numerical approaches, thus providing a test and validation for the conclusions presented in this work.

In particular, it is of interest to test the robustness of $l_0$. Provided that the simulation is in a quasi-steady state, the value of $l_0$ is largely determined by the surface density profile, $\Sigma$ [Eq.~(\ref{eq:sigma_balance}), with both $F(r)$ and $\nu(r)$ being the same for different numerical methods, within the errors of the hydrodynamical solver]. Since the diode boundary condition at $r = r_\mathrm{in}$ in our \textsc{pluto} simulations cannot capture the ``sloshing'' nature of the gas dynamics around eccentric binaries at that location, it is possible that even though $\dot{M}(r) = \mathrm{constant}$ in both sets of simulations, a slightly different value of $\Sigma$ at $r \approx r_\mathrm{in}$ might change the value of $l_0$. However, the qualitative agreement in $\dot{M}(t)$ between the two sets of simulations (see Figs.~$2$ and $5$ of Mu\~{n}oz \& Lai 2016) for different eccentricities and locations within the disc (except at $r = r_\mathrm{in}$) is encouraging. The fact that the value of $l_0$ calculated in our \textsc{pluto} simulations is independent of $r_\mathrm{in}$ (see Fig.~\ref{fig:ang_mom_balance}) also suggests the robustness of our $l_0$ values.

\section*{Acknowledgements}
This work has been supported in part by NASA grants NNX14AG94G and NNX14AP31G, and a Simons Fellowship from the Simons Foundation.

\appendix

\section{Angular Momentum Conservation in Circumbinary Discs}
\label{sec:ang_mom_cons}
The equation of motion for the gas in the disc is
\be
\frac{\partial\mathbf{u}}{\partial t} + \left(\mathbf{u}\cdot\mathbf{\nabla}\right)\mathbf{u} = -\frac{1}{\Sigma}\mathbf{\nabla} P - \mathbf{\nabla}\Phi + \mathbf{f}_\mathrm{visc},
\ee
where $\Phi$ is the gravitational potential of the binary and $\mathbf{f}_\mathrm{visc}$ is the viscous force per unit mass. Starting with the $\phi$ component,
\be
\frac{\partial u_\phi}{\partial t} + u_r \frac{\partial u_\phi}{\partial r} + \frac{u_\phi}{r} \left(u_r + \frac{\partial u_\phi}{\partial\phi}\right) = -\frac{1}{r\Sigma} \frac{\partial P}{\partial \phi} - \frac{1}{r}\frac{\partial\Phi}{\partial r} + f_{\mathrm{visc},\phi},
\ee
multiplying by $r$, and defining the angular momentum per unit mass $l = r u_\phi$, we have
\be
\label{eq:eom_l}
\frac{\partial l}{\partial t} + u_r\frac{\partial l}{\partial r} + \frac{u_\phi}{r}\frac{\partial l}{\partial\phi} = -\frac{1}{\Sigma}\frac{\partial P}{\partial \phi} - \frac{\partial \Phi}{\partial\phi} + t_z^\mathrm{visc},
\ee
where $t_z^\mathrm{visc} = rf_{\mathrm{visc},\phi}$ is the viscous torque per unit mass. Noting that left-hand side of this equation is equal to $\partial l/\partial t + (\mathbf{u}\cdot\mathbf{\nabla})l = \mathrm{d}l/\mathrm{d}t$, we see that this is the evolution equation for $l$.
Next, multiplying Eq.~(\ref{eq:eom_l}) by $r\Sigma$, making use of the identity
\be
\Sigma\frac{\partial l}{\partial t} = \frac{\partial}{\partial t}\left(\Sigma l\right) - l\frac{\partial\Sigma}{\partial t},
\ee
and the continuity equation,
\be
\frac{\partial\Sigma}{\partial t} = -\mathbf{\nabla}\cdot\left(\Sigma\mathbf{u}\right),
\ee
we arrive at
\be
\frac{\partial}{\partial t}\left(r\Sigma l\right) + \frac{\partial}{\partial r}\left(r\Sigma u_r l\right) + \frac{\partial}{\partial\phi}\left(\Sigma u_\phi l\right) = -r\frac{\partial P}{\partial\phi} - r\Sigma\frac{\partial\Phi}{\partial\phi} + r\Sigma t_z^\mathrm{visc}.
\ee
Integrating in $\phi$, we then have
\be
\label{eq:ang_mom_ring}
\frac{\partial}{\partial t}\left(\frac{\mathrm{d}J}{\mathrm{d}r}\right) = \frac{\partial\dot{J}_\mathrm{adv}}{\partial r} + \frac{\mathrm{d}T_\mathrm{grav}}{\mathrm{d}r} - \frac{\partial\dot{J}_\mathrm{visc}}{\partial r},
\ee
where we have defined
\be
\frac{\mathrm{d}J(r,t)}{\mathrm{d}r} = \oint r\Sigma l \mathrm{d}\phi,
\ee
\be
\dot{J}_\mathrm{adv}(r,t) = -\oint r\Sigma u_r l \mathrm{d}\phi,
\ee
\be
\frac{\mathrm{d}T_\mathrm{grav}(r,t)}{\mathrm{d}r} = -\oint r\Sigma \frac{\partial\Phi}{\partial\phi} \mathrm{d}\phi,
\ee
and
\be
\frac{\partial\dot{J}_\mathrm{visc}(r,t)}{\partial r} = -\oint r\Sigma t_z^\mathrm{visc} \mathrm{d}\phi,
\ee
which are the angular momentum per unit $r$, the inward flux of angular momentum due to advection, the gravitational torque per unit $r$, and the viscous torque per unit $r$. Eq.~(\ref{eq:ang_mom_ring}) describes the rate of change of angular momentum in a ring of unit width. In general, the viscous torque is given by
\be
\dot{J}_\mathrm{visc} = -\oint r^3\nu\Sigma \left[\frac{\partial}{\partial r}\left(\frac{u_\phi}{r}\right) + \frac{1}{r^2}\frac{\partial u_\mathrm{r}}{\partial \phi} \right] \mathrm{d}\phi.
\ee

Taking the time average of Eq.~(\ref{eq:ang_mom_ring}), and assuming a steady state ($\partial/\partial t \rightarrow 0$), we have
\be
\frac{\partial}{\partial r}\left(\langle\dot{J}_\mathrm{adv}\rangle - \langle\dot{J}_\mathrm{visc}\rangle\right) + \left\langle\frac{\mathrm{d}T_\mathrm{grav}}{\mathrm{d}r}\right\rangle = 0.
\ee
Finally, integrating radially from $r$ to $r_\mathrm{out}$, and noting that $\mathrm{d}T_\mathrm{grav}/\mathrm{d}r$ vanishes far from the binary, we find that
\be
\label{eq:ang_mom_constant}
\langle\dot{J}\rangle_r = \langle\dot{J}_\mathrm{adv}\rangle_{r_\mathrm{out}} - \langle\dot{J}_\mathrm{visc}\rangle_{r_\mathrm{out}}.
\ee
We have defined
\be
\label{eq:jdot_def}
\langle\dot{J}\rangle_r \equiv \langle\dot{J}_\mathrm{adv}\rangle_r - \langle\dot{J}_\mathrm{visc}\rangle_r - \langle T_\mathrm{grav}^{>r}\rangle,
\ee
where
\be
\langle T_\mathrm{grav}^{>r}\rangle \equiv \int_r^{r_\mathrm{out}}\left\langle\frac{\mathrm{d}T_\mathrm{grav}}{\mathrm{d}r}\right\rangle \mathrm{d}r.
\ee
Eq.~(\ref{eq:jdot_def}) gives the net angular momentum flux through the disc at radius $r$, including contributions from mass advection, viscous stress and gravitational torque. We also define its non-averaged analogue $\dot{J}$, by dropping the time averages. Eq.~(\ref{eq:ang_mom_constant}) indicates that $\langle\dot{J}\rangle$ is a constant across the disc.

Since $\dot{J}_\mathrm{visc} = 0$ at the (physical) inner edge of the disc, we find that $\langle\dot{J}\rangle_{r_\mathrm{in}} = \langle\dot{J}_\mathrm{adv}\rangle_{r_\mathrm{in}} - \langle T_\mathrm{grav}^\mathrm{tot}\rangle$ (where $\langle T_\mathrm{grav}^\mathrm{tot}\rangle$ is the gravitational torque exerted on the entire disc) is the net angular momentum transfer rate to the binary, which we denote $\langle\dot{J}_\mathrm{B}\rangle$. Therefore,
\be
\langle\dot{J}\rangle_r = \langle\dot{J}_\mathrm{B}\rangle
\ee
for all $r$. In a steady state, the mass accretion rate $\dot{M}$ is constant throughout the disc, therefore, we can write
\be
\dot{J}_\mathrm{B} = \dot{M}l_0.
\ee
The specific angular momentum $l_0$ represents the net angular momentum received by the binary per unit mass it accretes from the disc.

\end{document}